\documentclass[aps,prd,preprintnumbers,showpacs,showkeys,nofootinbib,
superscriptaddress,fleqn,floatfix,tightenlines,10pt]{revtex4-1}
\usepackage{amsmath,amsfonts,amssymb,amscd,amsxtra,amsthm}
\usepackage{graphicx}  
\usepackage{epstopdf}
\usepackage{dcolumn}  
\usepackage{bm}          
\usepackage{slashed}
\usepackage{cancel}
\usepackage{float}
\usepackage{mathtools}
\usepackage{amsbsy}
\usepackage{amstext}

\usepackage[utf8]{inputenc} 
\usepackage{booktabs} 
\usepackage[normalem]{ulem} 
\usepackage[dvipsnames]{xcolor} 
\usepackage{tabularx}
\usepackage{enumitem}  
\usepackage{array} 
\usepackage{slashed}
\usepackage{tikz}
\usepackage{float}
\usepackage{multirow}
\usepackage{cancel}
\renewcommand\sout{\bgroup \color{red} \ULdepth=-.5ex \ULset}

\makeatletter

\begin{document}  
\title{Electromagnetic multipole structure of a spin-one particle: Abel tomography case} 
\author{June-Young Kim}
\email[E-mail: ]{Jun-Young.Kim@ruhr-uni-bochum.de}
\affiliation{Institut f\"ur Theoretische Physik II, Ruhr-Universit\"at
  Bochum, D-44780 Bochum, Germany}

\date{\today}
\begin{abstract}
We investigate the three-dimensional~(3D) and two-dimensional~(2D) charge distributions of a spin-one particle in terms of the multipole expansion. On account of the geometrical difference between 2D and 3D spaces, projecting the 3D electric distribution to the 2D one in the Breit frame brings about the influence of the quadrupole distribution upon the monopole one. Thus, the 2D charge distribution becomes spin-dependent. This effect should be sorted out from the relativistic effects arising from the Lorentz boost. We first provide the connections between the 2D and 3D distributions in the Breit frame in terms of the angle-dependent Abel transformation. We then provide the differential equations that enable us to map 2D distributions in the Breit frame to those in the infinite momentum frame.
\end{abstract}
\maketitle

\section{Introduction}
The electromagnetic~(EM) form factors of a hadron have been considered  one of the important issues since measuring them in elastic-scattering experiments~\cite{Hofstadter:1956qs}. The EM form factors encode the information on the internal distributions of the charge and the magnetization. They have been simply obtained by the 3D Fourier transform~\cite{Ernst:1960zza, Sachs:1962zzc} of the EM form factors in the Breit frame~(BF). This interpretation is true only when the size of the system is much larger than the Compton wavelength. Since the Compton wavelength gives the minimum radius of a localized particle with a finite mass, the region smaller than the Compton wavelength brings about ambiguous relativistic effects. It means that non-relativistic treatment is valid for the particles such as atoms and nuclei. For example, in the case of the deuteron, the Compton wavelength is about $5\%$ of its charge radius. It indicates that the non-relativistic treatment of the deuteron is plausible. However, as the precision of the experiment gradually increased, a fully relativistic understanding of the EM distributions is needed instead of using the non-relativistic approximation. In addition, $\rho$-meson which is another spin-one particle is a fully relativistic particle because the Compton wavelength is about $30\sim40\%$ of the charge radius. It is thus important to explain the spin-one particle in a fully relativistic picture. This problem was first raised by Yennie \emph{et al.} many years ago~\cite{Yennie:1957} in the case of the nucleon. The ambiguous relativistic corrections were then removed by employing the two-dimensional~(2D) EM distributions in the infinite momentum frame~(IMF)~\cite{Burkardt:2000za, Burkardt:2002hr, Carlson:2007xd, Miller:2007uy, Miller:2010nz, Carlson:2009ovh, Alexandrou:2009hs}, since they are kinematically suppressed  in this frame. Recently, various prescriptions for this matter have been discussed in Ref.~\cite{Lorce:2020onh, Panteleeva:2021iip, Jaffe:2020ebz, Epelbaum:2022fjc, Lorce:2022jyi}.

Among those prescriptions, there was an attempt to grasp the internal structure of a hadron in terms of the Wigner distribution recently~\cite{Lorce:2018zpf, Lorce:2018egm}. At distance smaller than the Compton wavelength, one is able to interpret the EM distributions as quasi-probabilistic ones in the perspective of the phase-space Wigner distributions~\cite{Wigner:1932eb, Hillery:1983ms, Lee:1995}. This allows us to be free from ambiguous relativistic corrections. The price to pay is that we have a quasi-probabilistic distribution instead of having a strict probabilistic one. By adopting the phase-space Wigner distribution, Lorc\'{e} described nicely how the BF and IMF distributions can be naturally interpolated with each other for both the nucleon~\cite{Lorce:2020onh} and deuteron~\cite{Lorce:2022jyi}. This analysis will be kept to understand the BF static distributions in this paper. In addition to that approach, Panteleeva and Polyakov~\cite{Panteleeva:2021iip} have shown how the BF distributions can be mapped directly onto the IMF ones by using the Abel transformation~\cite{Abel} in the case of the nucleon mechanical properties. It was immediately extended to the energy-momentum tensor~\cite{Kim:2021jjf} and EM~\cite{Kim:2021kum} distributions. In fact, the application of the Abel tomography to the hadron structure has been already introduced in Ref.~\cite{Polyakov:2007rv, Moiseeva:2008qd}. Meanwhile, it was left behind as future work to study a hadron structure of a higher spin particle in the sense of tomography.

In the present work, we aim at investigating the 2D charge
distribution of a spin-one particle in the context of the Wigner sense and the Abel tomography. If a target is boosted to the IMF, the information on the 3D EM distribution gathers on the 2D plane via a line integral, so that they are subjected to both the geometrical and relativistic effects. However, in Ref.~\cite{Carlson:2009ovh} it was impossible to distinguish between relativistic and geometrical effects on the charge distribution. Thus, we will mainly focus on how the charge distribution is affected by those effects. In general, this charge distribution has been obtained by the 2D Fourier transformation of the helicity-amplitude form factors. However, it was rather difficult to grasp the physical meaning of each form factor, especially for higher-spin particles. We thus formulate the multipole distributions and form factors in the three different frames 3D BF, 2D BF, and 2D IMF instead of using the helicity-amplitude form factors. We will explicitly show how the projection from the 3D to 2D BF frames affects the monopole distribution in the presence of the quadrupole structure with the help of the angle-dependent Abel tomography. In fact, this projection results in the spin-dependent charge distribution in the 2D BF. This feature can be observed in the mass distribution and the mechanical properties of a higher spin particle~($S\geq1$). Note that the geometrical feature should be separated from the boost effects. After then, we take into account the Lorentz boost effects on the distribution via differential equations. In addition, it will be shown that the Lorentz boost is the other origin of the spin-dependent charge distribution.

We sketch this work as follows: In Sec. II, we present the formalism for the EM multipole form factors of a spin-one particle in three different frames 3D BF, 2D BF, and 2D IMF. In Sec. III, we briefly review how we define the EM multipole distributions in the Wigner sense. Then we present the EM multipole distributions in the three different frames 3D BF, 2D BF, and 2D IMF and relate them to each other in terms of the angle-dependent Abel transformation and differential equations. In Sec. III, we provide the numerical results for the deuteron charge distributions. Section V is devoted to the summary and conclusions of this work.
\section{Definition of the EM form factors
\label{sec:2}} 

The matrix element of the EM current operator $\hat{J}^{\mu}(r)=\sum_{f} q_{f} \bar{\psi}_{f}(r) \gamma^{\mu} {\psi}_{f}(r)$ with the flavor $f$ and the corresponding charge $q_{f}$ is parametrized in terms of the three different form factors $G_{1,2,3}$ for a spin-one particle as~\cite{Arnold:1979cg} (see also \cite{Cotogno:2019vjb} for a higher spin particle) 
\begin{align}
\langle  p', \lambda' | \hat{J}^{\mu}(0) | p, \lambda
  \rangle=  -2\left(\epsilon'^{*}\cdot \epsilon G_{1}(t) + 2 G_{3}(t) \frac{\epsilon'^{*}\cdot P \epsilon \cdot P}{m^{2}}\right) P^{\mu} + 2G_{2}(t) (\epsilon^{\mu} \epsilon'^{*} \cdot P + \epsilon'^{*\mu}\epsilon \cdot P),
  \label{eq:def_threeFF}
\end{align}
where $\lambda(\lambda')$ denotes the helicity of the initial (final) state of the spin-one particle. Note that both the initial and final momentum satisfy the on-mass-shell conditions $p^{2}=p'^{2}=m^{2}$. Here  the covariant normalization $\langle p', \lambda'| p, \lambda
\rangle = 2p^{0} (2\pi)^{3}\delta_{\lambda'  \lambda}
\delta^{(3)}(\bm{p}'-\bm{p})$ for the one-particle states is used, and we introduce the timelike average four-momentum $P=(p+p')/2$ and the spacelike four-momentum transfer $\Delta=p'-p$ with $\Delta^{2} = t$. The polarization vectors are defined as $\epsilon'_{\mu}=\epsilon'_{\mu}(p',\lambda')$, $\epsilon_{\mu}=\epsilon_{\mu}(p,\lambda)$ for brevety's sake. The explicit expression of the spin-one vector $\epsilon^{\mu}$ in any frame is given by
\begin{align}
\epsilon^{\mu}(p,\lambda) = \left( \frac{\bm{p}\cdot \hat{\bm{\epsilon}}_{\lambda}}{ m},\hat{\bm{\epsilon}}_{\lambda} + \frac{\bm{p}\cdot \hat{\bm{\epsilon}}_{\lambda}}{ m(m+p_{0})}\bm{p} \right),
\end{align}
where the spin-one polarization vector in the rest frame in the cartesian basis is given by
\begin{align}
\hat{\bm{\epsilon}}_{1}=(1,0,0), \ \ \ \hat{\bm{\epsilon}}_{2}=(0,1,0), \ \ \  \hat{\bm{\epsilon}}_{3}=(0,0,1).
\end{align}

\subsection{3D space}

Before discussing the multipole expansion of the EM form factors, we define the $n$-rank irreducible tensor and multipole operators. For a spin-one particle, it has a quadrupole structure. Thus, we first define the quadrupole operator $\hat{Q}^{ij}$(rank 2 tensor) in terms of the spin operator $\hat{S}^{i}$ as
\begin{align}
\hat{Q}^{ij} &= \frac{1}{2}\left( \hat{S}^{i}\hat{S}^{j} +\hat{S}^{j}\hat{S}^{i} -\frac{2}{3}S(S+1)\delta^{ij}\right),
\end{align}
with $i,j,k=1,2,3$. The operator is symmetrized and satisfy tracelessness~($\hat{Q}^{ii}=0$). The matrix element of the quadrupole operator is given by $\hat{Q}^{ij}_{\lambda' \lambda}$, and that of the spin operator can be expressed in terms of SU(2) Clebsch-Gordan coefficients in the spherical basis as
\begin{align}
\hat{S}^{a}_{\lambda'\lambda} = \sqrt{S(S+1)}C^{S \lambda'}_{S \lambda 1 a} \ \ \ \mathrm{with} \ \ \ (a=0,\pm1. \  \  \lambda,\lambda'=0, \cdot\cdot\cdot,\pm S).
\end{align}
Note that the 3D $n$-rank irreducible tensor in position space is defined as follows:
\begin{align}
Y_{0}(\Omega_{r})=1, \ \  Y^{i}_{1}(\Omega_{r})=\frac{r^{i}}{r},  \ \ Y^{ij}_{2}(\Omega_{r})=\frac{r^{i}r^{j}}{r^{2}}-\frac{1}{3}\delta^{ij}.
\label{eq:tensor}
\end{align}

We are now in a position to define the EM multipole form factors. In the BF, we set $\Delta^0=0$ and $\bm{P}=\bm{0}$, which means $p^0=p'^0=P^0$ and $\bm{p}' = -\bm{p}$.
The matrix element of the EM current for both the temporal and spatial components are then parametrized in terms of the multipole form factors:
\begin{align}
&\frac{\langle  p', \lambda' | \hat{J}^{0}(0) | p, \lambda
  \rangle}{2P_{0}}=  \delta_{\lambda'\lambda} G_{C}(t)  -2\tau \hat{Q}^{ij}_{\lambda' \lambda} Y^{ij}_{2}(\Omega_{q}) G_{Q}(t), \ \ \frac{\langle  p', \lambda' | \hat{J}^{i}(0) | p, \lambda
  \rangle}{2P_{0}}=  i \epsilon^{ijk} \hat{S}^{j}_{\lambda'\lambda} Y^{k}_{1}(\Omega_{q}) \sqrt{\tau}G_{M}(t),
\end{align}
where each term stands for the electric
$G_{C}(t)$, magnetic $G_{M}(t)$, and electric quadrupole $G_{Q}(t)$ form factors, respectively. They can be expressed as a linear combination of the three different form factors given in Eq.~\eqref{eq:def_threeFF}:
\begin{align}
G_{C}(t) = G_{1}(t) + \frac{2}{3}\tau G_{Q}(t), \ \ \ G_{M}(t) = G_{2}(t), \ \ \ G_{Q}(t) = G_{1}(t) -G_{2}(t) + (1+\tau) \tau G_{3}(t),
\end{align}
with $\tau=  \bm{\Delta}^{2}/4m^{2}$. The normalizations of the multipole form factors are defined as the charge $G_{C}(0)=C[e]$, the magnetic moment $G_{M}(0)=\mu [e/2m]$, and the electric quadrupole moment $G_{Q}(0)=Q [e/m^{2}]$. 

\subsection{2D space}
Recently, the elastic frame~(EF)~\cite{Lorce:2018egm, Lorce:2020onh} was introduced and applied to study how the hadronic matrix element~\eqref{eq:def_threeFF} is varied under the Lorentz boost. It was shown that this frame naturally interpolates between the 2D BF and 2D IMF for both the nucleon~\cite{Lorce:2020onh} and the deuteron~\cite{Lorce:2022jyi}. In addition, this frame also allows one to define a quasi-probabilistic distribution for a moving hadron in the Wigner sense. To trace down the origin of both the geometrical and relativistic effects, we first examine how the multipole structure of the EM matrix element is given in 2D space. If we restrict ourselves to the 2D space, we have to define the 2D $n$-rank irreducible tensor as follows:
\begin{align}
Y^{\mathrm{(2D)}}_{0}(\theta_{x_{\perp}})=1, \ \  Y^{\mathrm{(2D)}i}_{1}(\theta_{x_{\perp}})=\frac{x_{\perp}^{i}}{x_{\perp}},  \ \ Y^{\mathrm{(2D)}ij}_{2}(\theta_{x_{\perp}})=\frac{x_{\perp}^{i}x_{\perp}^{j}}{x_{\perp}^{2}}-\frac{1}{2}\delta^{ij}, \ \ \ (i,j=1,2)
\label{eq:tensor}
\end{align}
In the EF, the spacelike momentum transfer $\bm{\Delta}=(\bm{\Delta}_{\perp},0)$ lies in the transverse plane. The frame satisfy conditions, $\bm{P}=(\bm{0},P_{z})$, and $\Delta^0 = 0$.
If we take 2D BF, i.e., $P_{z}=0$, the matrix element of the EM current for both the temporal and spatial components are expressed as 
\begin{align}
&\frac{\langle  p', \lambda'| \hat{J}^{0}(0) | p, \lambda
  \rangle}{2P_{0}} \bigg{|}_{P_{z}\to0}=  \delta_{3\lambda} \delta_{\lambda'3} G_{C1}(t)+ \delta_{\sigma'\sigma}  G_{C2}(t)  -2\tau \hat{Q}^{ij}_{\lambda' \lambda} Y^{\mathrm{(2D)}ij}_{2}(\theta_{q}) G_{Q}(t), \cr
&\frac{\langle  p', \lambda' | \hat{J}^{i}(0) | p, \lambda
  \rangle}{2P_{0}} \bigg{|}_{P_{z}\to0}=  i \epsilon^{ijk} \hat{S}^{j}_{\lambda'\lambda} Y^{(\mathrm{2D})k}_{1}(\theta_{q}) \sqrt{\tau}G_{M}(t),
  \label{eq:EFFF}
\end{align}
where we introduce the 2D BF form factors
\begin{align}
G_{C1}(t) =  \left(G_{C}(t)- \frac{2}{3}\tau G_{Q}(t)\right), \ \ \ G_{C2}(t) = \left(G_{C}(t)+ \frac{1}{3}\tau G_{Q}(t)\right),
\end{align}
with $\lambda=1,2,3$ and $\sigma=1,2$. $G_{C1}(t)$ is the charge form factor when a particle spin is polarized along the $z$-axis, whereas the $G_{C2}(t)$ is that when the particle spin is transversely polarized  to the $z$-axis. The distinctive feature compared to the 3D BF results is that the charge distribution is not independent of the spin polarization anymore when the 3D charge distribution is projected to 2D space. It originates from the presence of the quadrupole structure and totally comes from the geometrical difference between the 2D and 3D spaces. This feature can be observed in both the charge and mass~\cite{Freese:2019bhb} distributions of a higher spin particle~$(S\geq1)$ such as the $\rho$ meson and the $\Delta$ baryon. Here, one must bear in mind that they are not relativistic effects.

To estimate the relativistic effects, we take the IMF~($P_{z}\to \infty$)~\cite{Lorce:2022jyi}. We then recover the expressions presented in light-cone quantization~\cite{Carlson:2009ovh}:
\begin{align}
\frac{\langle  p', \lambda'| \hat{J}^{0}(0) | p, \lambda
  \rangle}{2P_{0}} \bigg{|}_{P_{z}\to \infty}&=  \delta_{3\lambda} \delta_{\lambda'3} G^{\mathrm{IMF}}_{C1}(t)+ \delta_{\sigma'\sigma}  G^{\mathrm{IMF}}_{C2}(t) \cr
  &+i \epsilon^{3jk} \hat{S}^{j}_{\lambda'\lambda} Y^{(\mathrm{2D})k}_{1}(\theta_{q}) \sqrt{\tau}G^{\mathrm{IMF}}_{M}(t)  -2\tau \hat{Q}^{ij}_{\lambda' \lambda} Y^{\mathrm{(2D)}ij}_{2}(\theta_{q}) G^{\mathrm{IMF}}_{Q}(t),
\end{align}
where we introduce the multipole form factors
\begin{align}
&G^{\mathrm{IMF}}_{C1}(t) =  \left[G_{C}(t)- \frac{2}{3}\tau G_{Q}(t)\right]-2\tau G^\mathrm{IMF}_{W}(t), \ \ \ G^{\mathrm{IMF}}_{C2}(t) = \left[G_{C}(t)+ \frac{1}{3}\tau G_{Q}(t)\right] -\tau G^\mathrm{IMF}_{W}(t), \cr
&G^{\mathrm{IMF}}_{M}(t) =  -G_{M}(t)-2 G^\mathrm{IMF}_{W}(t), \ \ \ G^{\mathrm{IMF}}_{Q}(t) = G_{Q}(t) - G^\mathrm{IMF}_{W}(t),
\label{eq:IMF_FF}
\end{align}
with
\begin{align}
G^{\mathrm{IMF}}_{W}(t)=\frac{1}{1+\tau}\left( G_{C}(t)- G_{M}(t)+\frac{1}{3}\tau G_{Q}(t) \right).
\end{align}
It is now easy to grasp the meaning of each term in Eq.~\eqref{eq:IMF_FF}. In the case of the electric form factors $G^{\mathrm{IMF}}_{C1}$ and $G^{\mathrm{IMF}}_{C2}$, the first term inside the bracket is the charge form factor $G_{C}$ in the 3D BF whereas the second term inside the bracket originates from the geometrical difference between the 2D and 3D spaces. The last term $G_{W}$ is due to the Wigner spin rotation under the Lorentz boost. Interestingly, it was found that this Wigner rotation effect can be parametrized in terms of the single combination of the form factors~\cite{Lorce:2022jyi}. Similarly, the quadrupole form factor $G^{\mathrm{IMF}}_{Q}$ can be understood. It consists of the quadrupole form factor $G_{Q}$ in the 3D BF and the Wigner rotation effect $G_{W}$. On the other hand, the magnetization form factor $G^{\mathrm{IMF}}_{M}$ is solely due to  relativistic effects. As shown in Eq.~\eqref{eq:EFFF} there is no magnetization contribution to the temporal component of the EM current in the BF. However, when the system starts to be boosted, the Wigner spin rotation brings about the relativistic correction $G_{W}$. In addition to that, the mixture of the spatial and temporal components under the Lorentz boost results in the first term $G_{M}$. The normalizations of the IMF form factors can be found as follows:
\begin{align}
G^{\mathrm{IMF}}_{C1}(0)=G^{\mathrm{IMF}}_{C2}(0)=G_{C}(0), \ \ G^{\mathrm{IMF}}_{M}(0)=G_{M}(0)-2G_{C}(0), \ \ G^{\mathrm{IMF}}_{Q}(0)=G_{Q}(0)-G_{C}(0)+G_{M}(0). 
\end{align}
The given results are consistent with the those in Ref.~\cite{Carlson:2009ovh, Lorce:2022jyi}\footnote{One is able to find the connection between the multipole form factors and the helicity-amplitude form factors~\cite{Lorce:2022jyi}, such as $\mathcal{A}^{\mathrm{IMF}}_{00}=G^{\mathrm{IMF}}_{C1}, \mathcal{A}^{\mathrm{IMF}}_{11}=G^{\mathrm{IMF}}_{C2}, \mathcal{A}^{\mathrm{IMF}}_{01}=\sqrt{\frac{\tau}{2}}G^{\mathrm{IMF}}_{M}$, and  $\mathcal{A}^{\mathrm{IMF}}_{-11}=-\tau G^{\mathrm{IMF}}_{Q}.$}.

\section{Definition of the EM distributions
\label{sec:2}} 
While the 3D charge distribution of a spin-one particle cannot be interpreted as a probabilistic distribution because of the ambiguous relativistic corrections, it can be understood as a quasi-probabilistic distribution by means of the Wigner distribution. This quasi-probabilistic distribution conveys information on the internal structure of a hadron in a fully relativistic picture. The matrix element of the EM current for a physical state $|\psi \rangle$ can be defined as~\cite{Lorce:2020onh} 
\begin{align}
\langle \hat{J}^{\mu}(\bm{r}) \rangle= \int
  \frac{d^{3} \bm{P}}{(2\pi)^{3}}\int d^{3} \bm{R} \, W(\bm{R},\bm{P})
  \langle \hat{J}^{\mu}(\bm{r}) \rangle_{\bm{R},\bm{P}}, 
\label{eq:1}
\end{align}
where $W(\bm{R},\bm{P})$ stands for the Wigner distribution that is
given by 
\begin{align}
W(\bm{R},\bm{P}) &=\int \frac{d^{3} \bm{\Delta}}{(2\pi)^{3}}
   e^{-i\bm{\Delta}\cdot \bm{R}}   \tilde{\psi}^{*}\left(\bm{P} +
   \frac{\bm{\Delta}}{2}\right)   \tilde{\psi}\left(\bm{P} -
   \frac{\bm{\Delta}}{2}\right) \cr 
&=\int d^{3}\bm{z} \,e^{-i\bm{z}\cdot \bm{P}} {\psi}^{*}\left(\bm{R} -
  \frac{\bm{z}}{2}\right) {\psi}\left(\bm{R} +
  \frac{\bm{z}}{2}\right). 
\label{eq:Wig}
\end{align}
The average position $\bm{R}$ and momentum $\bm{P}$ are defined as
$\bm{R}=(\bm{r}'+\bm{r})/2$ and $\bm{P}=(\bm{p}'+\bm{p})/2$, respectively. 
$\bm{\Delta}=\bm{p}'-\bm{p}$ denotes the momentum transfer, which enables us to access the internal structure of a particle. The variable
$\bm{z}=\bm{r}'-\bm{r}$ represents the position separation between
the initial and final particles. 
The Wigner distribution contains information on the wave packet of a 
particle 
\begin{align}
 \psi(\bm{r}) = \langle \bm{r} |\psi \rangle  = \int \frac{d^3
  \bm{p}}{(2\pi^3)} e^{i\bm{p}\cdot \bm{r}} 
  \tilde{\psi}(\bm{p}), \ \ \  \tilde{\psi}(\bm{p}) =
  \frac{1}{\sqrt{2p^{0}}}\langle p| \psi \rangle, 
\end{align}
with the plane-wave states $|p\rangle$ and
  $|\bm{r}\rangle$ respectively normalized as  
 $\langle p'|p\rangle = 2 p^{0}(2\pi)^3
\delta^{(3)}(\bm{p}'-\bm{p})$ and $\langle \bm{r}'|\bm{r}\rangle
=  \delta^{(3)}(\bm{r}'-\bm{r})$. The position state
$|\bm{r}\rangle$ localized at $\bm{r}$ at time $t=0$ is defined
as a Fourier transform of the momentum eigenstate $|p\rangle$
\begin{align}
|\bm{r} \rangle = \int \frac{d^{3}\bm{p}}{(2\pi)^{3}\sqrt{2p^{0}}}
       e^{-i\bm{p}\cdot \bm{r}} | p \rangle. 
\end{align}
If we integrate
over the average position and momentum, then the probabilistic density
in either position or momentum space is recovered to be 
\begin{align}
  \int   \frac{d^{3}\bm{P}}{(2\pi)^{3}}\,W_{N}(\bm{R},\bm{P}) =|
  \psi_N\left(\bm{R} \right) |^{2},\;\;\;
\int d^{3}\bm{R}\,W_{N}(\bm{R},\bm{P}) =| \tilde{\psi}_N\left(\bm{P} \right)
  |^{2}.
\end{align}

Given $\bm{P}$ and $\bm{R}$, the matrix element $\langle
\hat{J}^{\mu} (\bm{r})\rangle_{\bm{R},\bm{P}}$ conveys information on
the internal structure of the particle localized around the average
position $\bm{R}$ and average momentum $\bm{P}$. 
This can be expressed as the 3D Fourier transform of the matrix
element $\langle  p', \lambda' | \hat{J}^{\mu}(0) | p, \lambda
\rangle$:   
\begin{align}
\langle \hat{J}^{\mu}(\bm{r}) \rangle_{\bm{R},\bm{P}}= \langle
  \hat{J}^{\mu}(0) \rangle_{-\bm{x},\bm{P}} = \int
  \frac{d^{3}\bm{\Delta}}{(2\pi)^{3}} e^{-i\bm{x} \cdot \bm{\Delta} }
  \frac{1}{\sqrt{2p^{0}}\sqrt{2p'^{0}}}\langle  p', \lambda' |
  \hat{J}^{\mu}(0) | p, \lambda \rangle, 
  \label{eq:5_1}
\end{align} 
with the shifted position vector
$\bm{x}=\bm{r}-\bm{R}$.

\subsection{3D Breit frame}
In the BF, we have $\Delta^0=0$ and $\bm{P}=\bm{0}$. Having integrated over $\bm{P}$ of Eq.~\eqref{eq:1}, we find that the part of the wave packet can be factorized. Thus, the target in the BF is understood as a localized state around $\bm{R}$ from the Wigner perspective. In this frame, Eq.~\eqref{eq:5_1} is reduced to  
\begin{align}
J^{0}_{\mathrm{BF}}(\bm{x},\lambda', \lambda):=\langle \hat{J}^{0}(0) \rangle_{-\bm{x},\bm{0}} 
=  \int
  \frac{d^{3}\bm{\Delta}}{(2\pi)^{3}} e^{-i\bm{x} \cdot \bm{\Delta} }
  \frac{1}{2P^{0}}\langle  p', \lambda' |
  \hat{J}^{0}(0) | p, \lambda \rangle. 
\label{eq:5}
\end{align} 
From now on we use $\bm{r}$ instead of $\bm{x}$. We introduce the temporal component of the EM distributions in terms of the multipole expansion as follows:
\begin{align}
J^{0}_{\mathrm{BF}}(\bm{r},\lambda', \lambda) = \rho_{C}(r) \delta_{\lambda'\lambda} + \rho_{Q}(r) \hat{Q}^{ij}_{\lambda'\lambda} Y^{ij}(\Omega_{r}),
\label{eq:5_multipole}
\end{align}
where the EM distributions are given in terms of the EM multipole form factors by
\begin{align}
\rho_{C,M}(r) = \tilde{G}_{C,M}(r), \ \ \rho_{Q}(r) = \frac{r}{2m^{2}}\frac{d}{dr} \frac{1}{r} \frac{d}{dr} \tilde{G}_{Q}(r), \ \ \tilde{G}_{C,M,Q}(r) = \int \frac{d^{3}\bm{\Delta}}{(2\pi)^{3}} e^{-i\bm{r} \cdot \bm{\Delta} }
  G_{C,M,Q}(\bm{\Delta}^{2}).
\end{align}
Note that the magnetization distribution is defined as $\bm{J}=\bm{\nabla} \times \bm{M}$. The EM multipole form factors $G_{C,M,Q}(t)$ can be also expressed in terms of the EM distributions $\rho_{C,M,Q}(r)$ in coordinate space:
\begin{align}
G_{C,M}(t) = \int d^{3}r \, j_{0}(r\sqrt{-t} ) \rho_{C,M}(r), \ \ \ G_{Q}(t) =-2m^{2} \int d^{3}r \, \frac{j_{2}(r\sqrt{-t} )}{t} \rho_{Q}(r).
\end{align}
At the zero momentum transfer $t=0$, the normalizations of the form factors are expressed as the integrals of the EM distributions over position $r$:
\begin{align}
&G_{C,M}(0) =\int d^{3}r \, \rho_{C,M}(r), \ \ \ G_{Q}(0) =\frac{2}{15}m^{2}\int d^{3}r \, r^{2}\rho_{Q}(r).
\end{align}
In addition, the charge and magnetic radii are defined as the slope of their multipole form factors $G_{C,M}$:
\begin{align}
\langle r^{2} \rangle_{C,M} = \frac{\int d^{3}r \, r^{2} \rho_{C,M}(r)}{\int d^{3}r \, \rho_{C,M}(r)} = \frac{6}{G_{C,M}(0)}\frac{d G_{C,M}(t)}{dt}\bigg{|}_{t=0}.
\end{align}

\subsection{2D Breit frame}
The EF distributions depend on the impact parameter $x_{\perp}$
($\bm{r}=(\bm{x}_{\perp}, \, x_{z})$) and momentum $\bm{P}=(\bm{0},P_{z})$ where a spin-one particle moves along the $z$-direction without loss of generality. In this frame, Eq.~\eqref{eq:5_1} is reduced to  
\begin{align}
J^{0}_{\mathrm{EF}}(\bm{x}_{\perp},P_{z},\lambda', \lambda):= \int dx_{z} \langle \hat{J}^{0}(0) \rangle_{-\bm{r},\bm{0}} 
=  \int
  \frac{d^{2}\bm{\Delta}_{\perp}}{(2\pi)^{2}} e^{-i\bm{x}_{\perp} \cdot \bm{\Delta}_{\perp} }
  \frac{1}{2P^{0}}\langle  p', \lambda' |
  \hat{J}^{0}(0) | p, \lambda \rangle \bigg{|}_{\Delta_{z}=0}.
\label{eq:6}
\end{align} 
Before investigating the IMF distributions, one should separate the geometrical contributions from the relativistic ones first. We thus examine the distributions in the 2D BF by taking $P_{z}\to0$. The temporal component of the EM current in the 2D EF are given by
\begin{align}
 J^{0}_{\mathrm{EF}}(\bm{x}_{\perp},0,\lambda',
  \lambda)&=  \delta_{3\lambda} \delta_{\lambda'3} \rho^{\mathrm{(2D)}}_{C1}(x_{\perp})+ \delta_{\sigma'\sigma}  \rho^{\mathrm{(2D)}}_{C2}(x_{\perp})+ \hat{Q}^{ij}_{\lambda' \lambda} Y^{(\mathrm{2D})ij}_{2}(\theta_{x_{\perp}}) \rho^{\mathrm{(2D)}}_{Q}(x_{\perp}),
\label{eq:j0}
\end{align}
where individual distributions are given in terms of the EM multipole form factors by
\begin{align}
&\rho^{\mathrm{(2D)}}_{C1,C2,M}(x_{\perp}) = \tilde{G}^{\mathrm{(2D)}}_{C1,C2,M1}(x_{\perp}), \ \ \ \rho^{\mathrm{(2D)}}_{Q}(x_{\perp}) = \frac{x_{\perp}}{2m^{2}}\frac{d}{dx_{\perp}} \frac{1}{x_{\perp}} \frac{d}{dx_{\perp}} \tilde{G}^{\mathrm{(2D)}}_{Q}(x_{\perp}), \cr
&\tilde{G}^{\mathrm{(2D)}}_{C1,C2,M,Q}(x_{\perp}) = \int \frac{d^{2}\bm{\Delta}_{\perp}}{(2\pi)^{2}} e^{-i\bm{x}_{\perp} \cdot \bm{\Delta}_{\perp} }
  G_{C,M,Q}(\bm{\Delta}^{2}_{\perp}).
\end{align}
Since there is no magnetization contribution to the temporal component of the EM current without the Lorentz boost, the magnetization distribution is defined through the spatial component of the EM current as usual. As pointed out in Eq.~\eqref{eq:EFFF}, the charge distribution split into the $\rho^{\mathrm{(2D)}}_{C1}$ and $\rho^{\mathrm{(2D)}}_{C2}$ due to the presence of the quadrupole structure. It will be explicitly verified in terms of the EM distributions in the next subsection. The EM multipole form factors $G_{C1,C2,M,Q}$ can be also expressed in terms of the EM distributions $\rho^{\mathrm{(2D)}}_{C1,C2,M,Q}$ in coordinate space:
\begin{align}
&G_{C1,C2,M}(t) = \int d^{2}x_{\perp} J_{0}(x_{\perp}\sqrt{-t} ) \rho^{\mathrm{(2D)}}_{C1,C2,M}(x_{\perp}), \ \ \ G_{Q}(t) =-2m^{2} \int d^{2}x_{\perp} \frac{J_{2}(x_{\perp} \sqrt{-t} )}{t} \rho^{\mathrm{(2D)}}_{Q}(x_{\perp}).
\end{align}
At the zero momentum transfer $t=0$, the normalizations of the 2D BF form factors are found to be
\begin{align}
&G_{C1,C2,M}(0) =\int d^{2}x_{\perp} \, \rho^{\mathrm{(2D)}}_{C1,C2,M}(x_{\perp}), \ \ \ G_{Q}(0) =\frac{m^{2}}{4}\int d^{2}x_{\perp} \, x_{\perp}^{2}\rho^{\mathrm{(2D)}}_{Q}(x_{\perp}),
\end{align}
In addition, the charge and magnetic radii are defined as the slope of their multipole form factors $G_{C1,C2,M}$:
\begin{align}
\langle x^{2}_{\perp} \rangle^{\mathrm{(2D)}}_{C1,C2,M} = \frac{\int d^{2}x_{\perp} \, x_{\perp}^{2} \rho^{\mathrm{(2D)}}_{C1,C2,M}(x_{\perp})}{\int d^{2}x_{\perp} \, \rho^{\mathrm{(2D)}}_{C1,C2,M}(x_{\perp})} = \frac{4}{G_{C1,C2,M}(0)}\frac{d G_{C1,C2,M}(t)}{dt}\bigg{|}_{t=0}.
\end{align}

\subsection{Beyond Abel Tomography}
In the various Refs.~\cite{Panteleeva:2021iip, Kim:2021jjf, Kim:2021kum}, it was examined that the spherical-symmetric 3D distributions for the nucleon can be directly mapped to 2D ones via Abel transformation. The Abel transformation and its inverse transformation are defined as
\begin{align}
A[g](x_{\perp}) =\mathcal{G}(x_{\perp}) = \int^{\infty}_{x_{\perp}}
  \frac{dr}{r} \frac{g(r)}{\sqrt{r^{2}-b^{2}}}, \ \ \  
g(r) =  - \frac{2}{\pi} r^{2} \int^{\infty}_{r} d x_{\perp}
  \frac{d\mathcal{G}(x_{\perp})}{dx_{\perp}} 
  \frac{g(r)}{\sqrt{x_{\perp}^{2}-r^{2}}}.   
\label{eq:Able}
\end{align}
Thus, $A[g](b):= \mathcal{G}(b)$ is called the Abel image of the
function $g(r)$. In addition, a useful relation for the Mellin moments of the Abel images can be obtained as~\cite{Panteleeva:2021iip}:
\begin{align}
& \int^{\infty}_{0} b^{N}   A[g](b)\,db=\frac{\sqrt{\pi}}{2}
  \frac{\Gamma\left(\frac{N+1}{2}\right)}{ 
\Gamma\left(\frac{N+2}{2}\right)} 
  \int^{\infty}_{0}  \, r^{N-1} g(r)  dr.
\label{eq:Mel}
\end{align}
For example, if there is no higher multipole distribution in the BF, the Abel image of the monopole charge distribution is found to be
\begin{align}
\int dx_{z} \langle \hat{J}^{0}(0) \rangle_{-\bm{r},\bm{0}} = \int dx_{z} \, \rho_{C}(r) \delta_{\lambda' \lambda} =\rho^{\mathrm{(2D)}}_{C}(x_{\perp}) \delta_{\lambda' \lambda}, \ \ \ \rho^{\mathrm{(2D)}}_{C}(x_{\perp})=\int^{\infty}_{x_{\perp}} dr \frac{2r \rho_{C}(r)}{\sqrt{r^{2}-x^{2}_{\perp}}}.
\label{eq:spherical_line}
\end{align}
It indeed works in the case of the nucleon and the pion. However, as we pointed out in the previous section, mapping the 3D charge distribution to 2D one in the presence of the quadrupole structure brings about additional contributions. To investigate their impact, we should employ an angle-dependent Abel transformation. The concept of this transformation is collecting all the angle-dependent Abel images in 3D space and reconstructing them in the 2D space. In our case, we need to integrate $\rho(r) Y^{ij}(\Omega_{r})$ over the $z$-axis for each 3D angle. Of course, one can postulate that the 3D and 2D distributions can be given by Eqs.~\eqref{eq:5_multipole} and~\eqref{eq:j0} in terms of the multipole expansion. Then, the angle-independent distributions $\rho(r)$ and $\rho^{\mathrm{(2D)}}(x_{\perp})$ can be connected through the Abel transformation, which is a method having been done in the following Refs.~\cite{Panteleeva:2021iip, Kim:2021jjf, Kim:2021kum}. However, one of the direct ways to relate them is to carry out the integral of the 3D BF in Eq.~\eqref{eq:5_multipole} over $z$. We found that the angle-dependent Abel transformation can be analytically implemented. It is derived as:
\begin{align}
 \int dx_{z} \, \rho_{Q}(r) Y^{ij}(\Omega_{r}) \hat{Q}^{ij}_{\lambda' \lambda} =\rho^{\mathrm{(2D)}}_{Q}(x_{\perp}) Y^{\mathrm{(2D)}ij}_{2}(\theta_{x_{\perp}}) \hat{Q}^{ij}_{\lambda' \lambda} +\Delta_{Q}(x_{\perp}) \left( -\frac{1}{3}\delta_{\sigma'\sigma} + \frac{2}{3}\delta_{\lambda' 3}\delta_{3\lambda} \right), 
\end{align}
with
\begin{align}
\rho^{\mathrm{(2D)}}_{Q}(x_{\perp})=\int^{\infty}_{x_{\perp}} dr \frac{2x^{2}_{\perp} \rho_{Q}(r)}{r\sqrt{r^{2}-x^{2}_{\perp}}}, \ \ \ \Delta_{Q}(x_{\perp})=\int^{\infty}_{x_{\perp}} dr \frac{(3x^{2}_{\perp}-2r^{2}) \rho_{Q}(r)}{r\sqrt{r^{2}-x^{2}_{\perp}}}.
\end{align}
Under this transformation, we observe that the $2$-rank irreducible tensor in the 3D space is reduced to the $2$-rank irreducible tensor in the 2D space and a part of the diagonal contributions leak out to the $0$-rank irreducible tensor in the 2D space. This \emph{induced monopole}  distribution $\Delta_{Q}$ is responsible for the splitting of the charge distributions with the longitudinally and transversely polarized spins. Since the energy-momentum tensor distributions for a higher-spin particle possess an intricate structure in comparison with the EM distributions, this geometrical understanding is indeed important.

By considering the above relations, we are now able to provide the explicit connections between the 2D and 3D BF distributions in terms of Abel transformations as follows:
\begin{align}
&\rho^{\mathrm{(2D)}}_{C1}(x_{\perp})+2\rho^{\mathrm{(2D)}}_{C2}(x_{\perp})  = 6 \int^{\infty}_{x_{\perp}} \frac{r dr}{\sqrt{r^{2}-x^{2}_{\perp}}} \rho_{C}(r) = 3\rho^{\mathrm{(2D)}}_{C}(x_{\perp}),  \cr 
&\Delta_{Q}(x_{\perp}):=\rho^{\mathrm{(2D)}}_{C1}(x_{\perp})-\rho^{\mathrm{(2D)}}_{C2}(x_{\perp})  = \frac{\partial^{2}_{(2D)}}{4m^{2}} \tilde{G}^{\mathrm{(2D)}}_{Q}(x_{\perp}), \ \ \ \rho^{\mathrm{(2D)}}_{M}(x_{\perp})  = 2 \int^{\infty}_{x_{\perp}} \frac{r dr}{\sqrt{r^{2}-x^{2}_{\perp}}} \rho_{M}(r).
\end{align}
From Eq.~\eqref{eq:Mel}, the obvious relations between the 2D and 3D distributions can be found to be
\begin{align}
&G_{C}(0)=\int d^{3}r \, \rho_{C}(r)= \int d^{2}x_{\perp} \, \rho^{\mathrm{(2D)}}_{C1}(x_{\perp})= \int d^{2}x_{\perp} \, \rho^{\mathrm{(2D)}}_{C2}(x_{\perp}), \cr
&G_{M}(0)=\int d^{3}r \, \rho_{M}(r)= \int d^{2}x_{\perp} \, \rho^{\mathrm{(2D)}}_{M}(x_{\perp}).
\end{align}
In addition, the charge radii and the quadrupole moments between the 2D and 3D distributions can be related as follows:
\begin{align}
&\langle x^{2}_{\perp} \rangle^{\mathrm{(2D)}}_{C1} G_{C}(0)  = \frac{2}{3}Q + \frac{2}{3}\langle r^{2} \rangle_{C} G_{C}(0), \ \ \ \langle x^{2}_{\perp} \rangle^{\mathrm{(2D)}}_{C2}G_{C}(0)  = -\frac{1}{3}Q + \frac{2}{3}\langle r^{2} \rangle_{C} G_{C}(0), \ \ \ \langle x^{2}_{\perp} \rangle^{\mathrm{(2D)}}_{M}  = \frac{2}{3}\langle r^{2} \rangle_{M}, \cr
&G_{Q}(0)=\frac{2}{15}m^{2}\int d^{3}r \, r^{2} \rho_{Q}(r)= \frac{m^{2}}{4}\int d^{2}x_{\perp} \, x_{\perp}^{2} \rho^{\mathrm{(2D)}}_{Q}(x_{\perp}).
\end{align}

\subsection{IMF}
We have obtained the 2D BF distributions through the angle-dependent Abel transformation, and they can be associated with the 2D IMF distributions via a specific differential equation. In the IMF, i.e., $P_{z}\to \infty$, we are able to write down the temporal component of the EM current in terms of the multipole expansion as follows:
\begin{align}
 J^{0}_{\mathrm{EF}}(\bm{x}_{\perp},\infty,\lambda',
  \lambda)&=  \delta_{3\lambda} \delta_{\lambda'3} \rho^{\mathrm{IMF}}_{C1}(x_{\perp})+ \delta_{\sigma'\sigma}  \rho^{\mathrm{IMF}}_{C2}(x_{\perp}) \cr
  &+i \epsilon^{3jk} \hat{S}^{j}_{\lambda' \lambda} Y_{1}^{\mathrm{(2D)}k}(\theta_{x_{\perp}}) \rho^{\mathrm{IMF}}_{M}(x_{\perp})+ \hat{Q}^{ij}_{\lambda' \lambda} Y^{\mathrm{(2D)}ij}_{2}(\theta_{x_{\perp}}) \rho^{\mathrm{IMF}}_{Q}(x_{\perp}).
\label{eq:jIMF}
\end{align}
Note that for simplicity we drop out the magnetization contribution in this work. The 2D IMF distributions are given in terms of the multipole EM form factors by
\begin{align}
&\rho^{\mathrm{IMF}}_{C1,C2}(x_{\perp}) = \tilde{G}^{\mathrm{IMF}}_{C1,C2}(x_{\perp}), \ \ \ \rho^{\mathrm{IMF}}_{Q}(x_{\perp}) = \frac{x_{\perp}}{2m^{2}}\frac{d}{dx_{\perp}} \frac{1}{x_{\perp}} \frac{d}{dx_{\perp}} \tilde{G}^{\mathrm{IMF}}_{Q}(x_{\perp}), \cr
&\tilde{G}^{\mathrm{IMF}}_{C1,C2,Q,W}(x_{\perp}) = \int \frac{d^{2}\bm{\Delta}_{\perp}}{(2\pi)^{2}} e^{-i\bm{x}_{\perp} \cdot \bm{\Delta}_{\perp} }
  G^{\mathrm{IMF}}_{C1,C2,Q,W}(\bm{\Delta}^{2}_{\perp}).
\end{align}
The 2D IMF distributions can be expressed in terms of 2D BF distributions through the given differential equations
\begin{align}
&\rho^{\mathrm{IMF}}_{C1}(x_{\perp}) = \rho^{\mathrm{(2D)}}_{C}(x_{\perp}) + \frac{2}{3} \Delta_{Q}(x_{\perp})+ 2\hat{\tau}\tilde{G}^{\mathrm{(2D)}}_{W}(x_{\perp}), \cr
&\rho^{\mathrm{IMF}}_{C2}(x_{\perp}) =  \rho^{\mathrm{(2D)}}_{C}(x_{\perp}) - \frac{1}{3}\Delta_{Q}(x_{\perp})+ \hat{\tau}\tilde{G}^{\mathrm{(2D)}}_{W}(x_{\perp}), \cr
&\rho^{\mathrm{IMF}}_{Q}(x_{\perp}) =\rho^{\mathrm{(2D)}}_{Q}(x_{\perp}) -\frac{1}{2m^{2}} x_{\perp} \frac{d}{dx_{\perp}} \frac{1}{x_{\perp}} \frac{d}{dx_{\perp}}   \tilde{G}^{\mathrm{(2D)}}_{W}(x_{\perp}).
\label{eq:decom}
\end{align}
with the dimensionless Laplacian $\hat{\tau}:=\frac{\partial^{2}_{(2D)}}{4m^{2}}$. One might notice that we encounter the notorious differential equations, since $\tilde{G}_{W}$ possesses the infinite order of the derivatives as pointed out in Ref.~\cite{Kim:2021kum, Freese:2021mzg}. However, we are still able to find their moments from Eq.~\eqref{eq:Mel} and truncate the differential equation up to a certain order, which may be a plausible approximation if a spin-one particle is a sufficiently heavy object. Here we provide the charge radii and the quadrupole moment in the IMF as follows:
\begin{align}
&\langle x^{2}_{\perp} \rangle^{\mathrm{IMF}}_{C1} G_{C}(0)  = \left(\frac{2}{3} \langle r^{2} \rangle_{C} G_{C}(0) + \frac{2}{3m^{2}} G_{Q}(0)  \right) + \frac{2}{m^{2}}G_{C}(0) - \frac{2}{m^{2}}G_{M}(0),  \cr
&\langle x^{2}_{\perp} \rangle^{\mathrm{IMF}}_{C2} G_{C}(0)  = \left(\frac{2}{3} \langle r^{2} \rangle_{C} G_{C}(0) - \frac{1}{3m^{2}} G_{Q}(0)  \right) + \frac{1}{m^{2}}G_{C}(0) - \frac{1}{m^{2}}G_{M}(0),  \cr
& Q^{\mathrm{IMF}} = \frac{1}{4} \int d^{2} x_{\perp} \, x^{2}_{\perp} \rho^{\mathrm{IMF}}_{Q}(x^{2}_{\perp}) = \bigg{[}-G_{C}(0)+G_{M}(0)+G_{Q}(0)\bigg{]} \frac{1}{m^{2}}.
\end{align}

\section{Deuteron 2D Charge distributions}
To verify the formalism constructed in the previous section we present and discuss the numerical results of the charge distributions of the deuteron in the three different frames, focusing on the origin of their spin-polarization dependences. To estimate them, we take the empirical parametrization of the EM form factors of the deuteron proposed in Ref.~\cite{JLABt20:2000qyq}. We first present how the charge distribution in the 3D BF changes into that in the 2D BF under the projection or line integral over the $z$-axis. 
\begin{figure}
\includegraphics[scale=0.275]{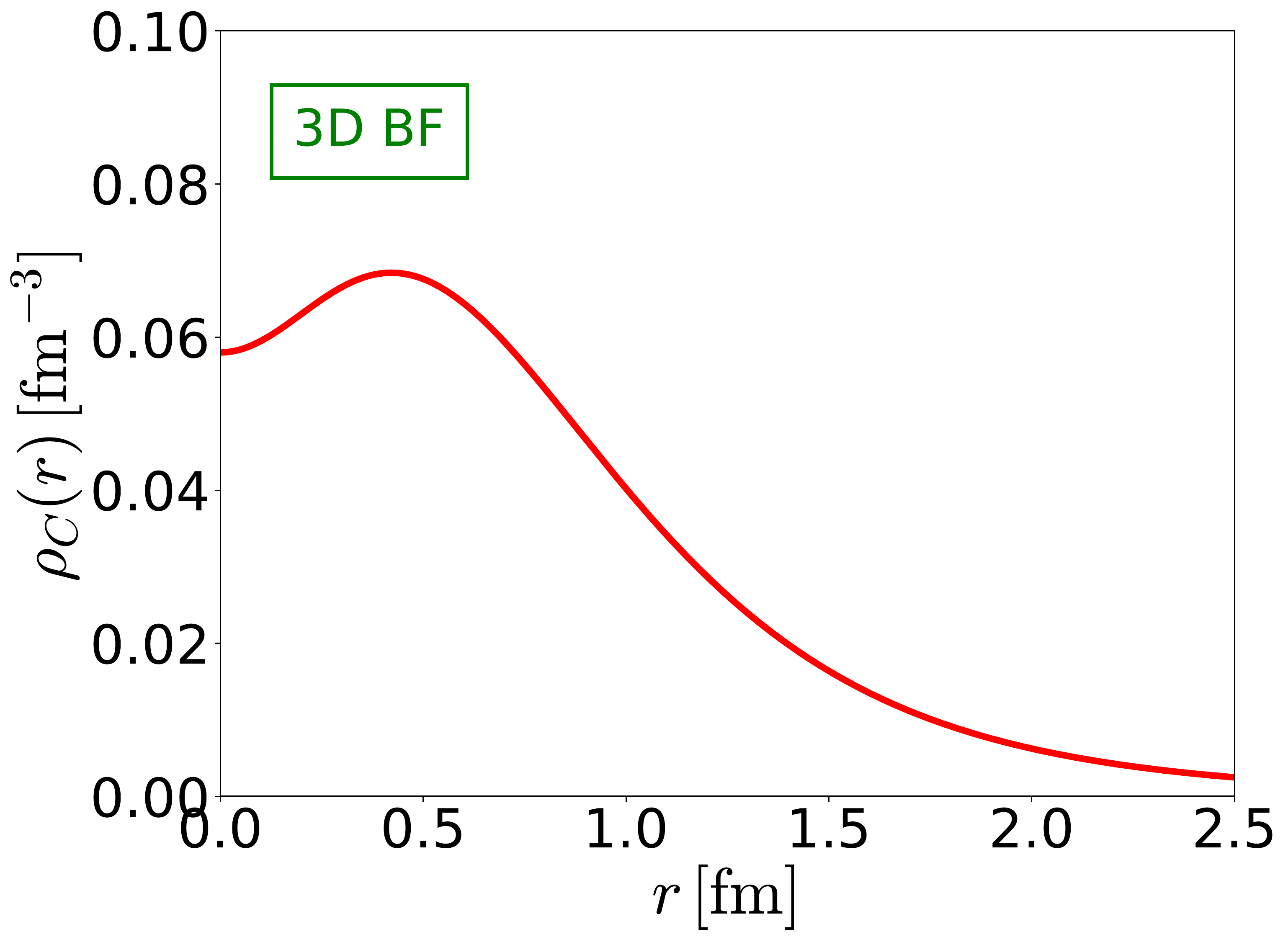} \hspace{1cm}
\includegraphics[scale=0.75]{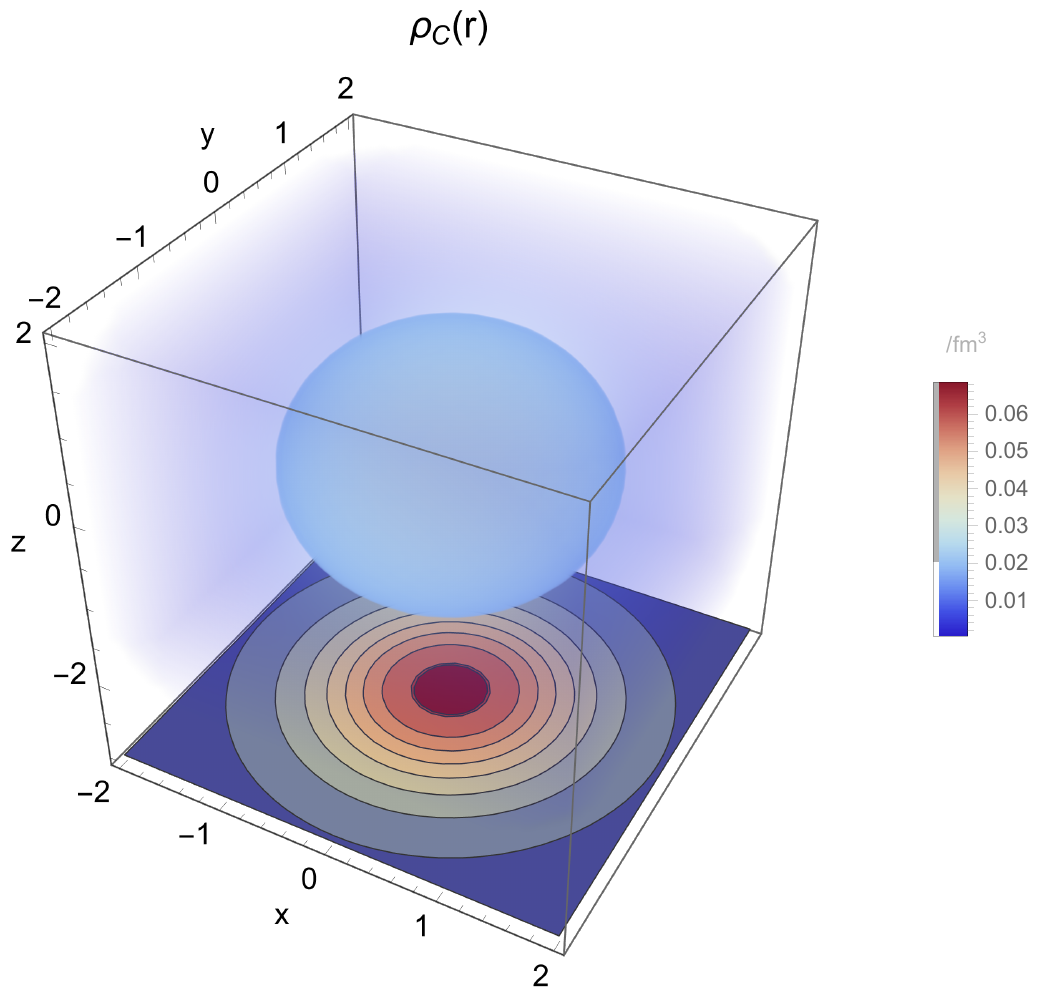}
\caption{In the left panel, we depict the 3D BF charge distribution of the deuteron as a function of $r$, and it is visualized in the 3D space in the right panel. We also present the 2D BF charge distribution on the bottom plane by integrating the 3D BF distribution over the $z$-axis.}
\label{fig:1}
\end{figure}
We draw the 3D BF charge distribution in the left panel of Fig.~\ref{fig:1}. It has a hole in the center, which is a typically well-known shape of the deuteron charge distribution. If we integrate it over the $z$-axis, all of the information on the 3D distribution is gathered on the 2D plane. It can be performed by the Abel transformation as shown in Eq.~\eqref{eq:spherical_line}. However, it is true only when a spherical-symmetric hadron is considered. As shown in Eq.~\eqref{eq:5_multipole}, the situation gets more complicated for the deuteron, since it has a quadrupole structure.

To carry out the angle-dependent Abel transformation, we first slice the quadrupole distribution $\rho_{Q}(r) Y_{2}^{ij}(\Omega_{r})$ with respect to each 3D angle $\Omega_{r}$. We then perform the Abel transformations for each 3D angle and obtain the corresponding Abel images. After then, we reconstruct them on a 2D plane. They are explicitly given by
\begin{align}
\int dz \, \rho_{Q}(r) Y_{2}^{ij}(\Omega_{r}) &= \rho^{\mathrm{(2D)}}_{Q}(x_{\perp}) \left(\begin{array}{c c c} \frac{1}{2}\cos{2\theta_{x_{\perp}}} & \cos{2\theta_{x_{\perp}}}\sin{\theta_{x_{\perp}}} & 0  \\ \cos{2\theta_{x_{\perp}}}\sin{\theta_{x_{\perp}}} & \frac{1}{2}\sin{2\theta_{x_{\perp}}} & 0  \\ 0 & 0 & 0 \end{array} \right)^{ij} + \Delta_{Q}(x_{\perp}) \left(\begin{array}{c c c} \frac{1}{3} & 0 & 0  \\ 0 & \frac{1}{3} & 0  \\ 0 & 0 & -\frac{2}{3} \end{array} \right)^{ij}.
\label{eq:reduce_ten}
\end{align}
As presented in Eq.~\eqref{eq:reduce_ten}, the 3D 2-rank irreducible tensor is reduced to the 2D $2$- and $0$-rank irreducible tensors. Since the off-diagonal components $(i=1, j=3)$, $(i=2, j=3)$, and $(i\leftrightarrow j)$ are proportional to the integral variable $z$, they are odd functions with respect to the plane $z=0$. Thus, the corresponding components vanish. On the other hand, the components $(i=1, j=2)$ and $(i=2, j=1)$ have pure quadrupole structures. While the diagonal component $(i=3, j=3)$ possesses monopole structure, the components $(i=1, j=1)$ and $(i=2, j=2)$ have both the monopole and quadrupole structures. It means that a part of the quadrupole distributions flows into the monopole one which is  named \emph{induced monopole} distribution. Interestingly, it differently contributes to the charge distribution according to the spin-polarization of the deuteron.
\begin{figure}
\includegraphics[scale=0.545]{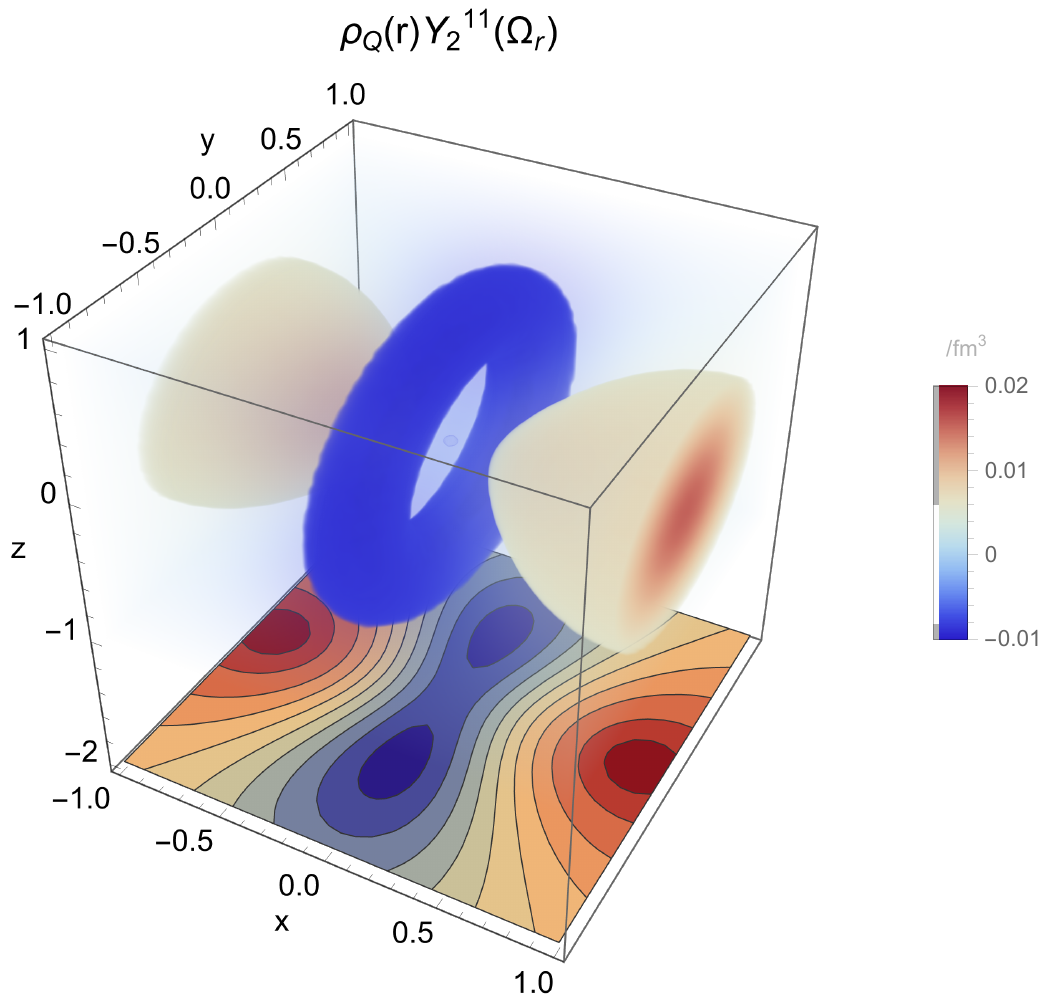}
\includegraphics[scale=0.545]{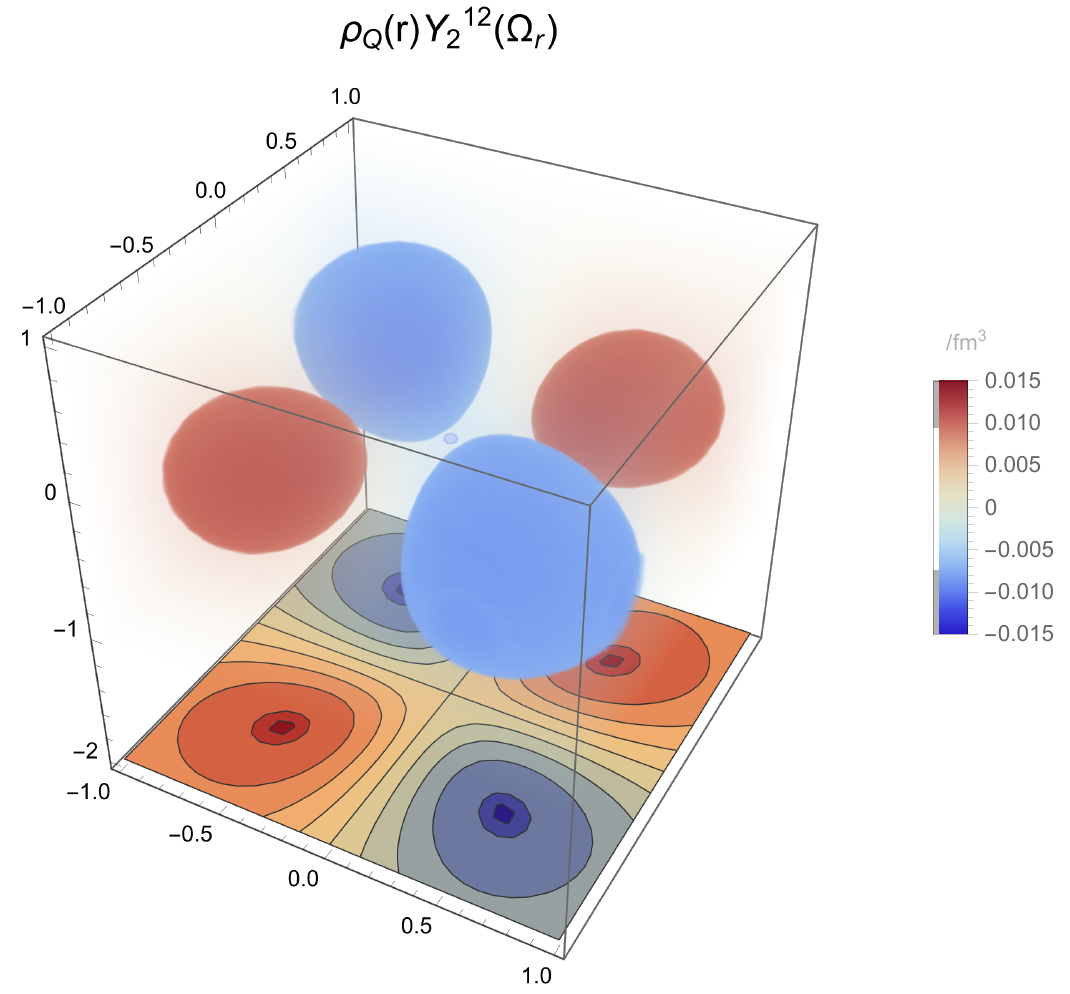}
\includegraphics[scale=0.545]{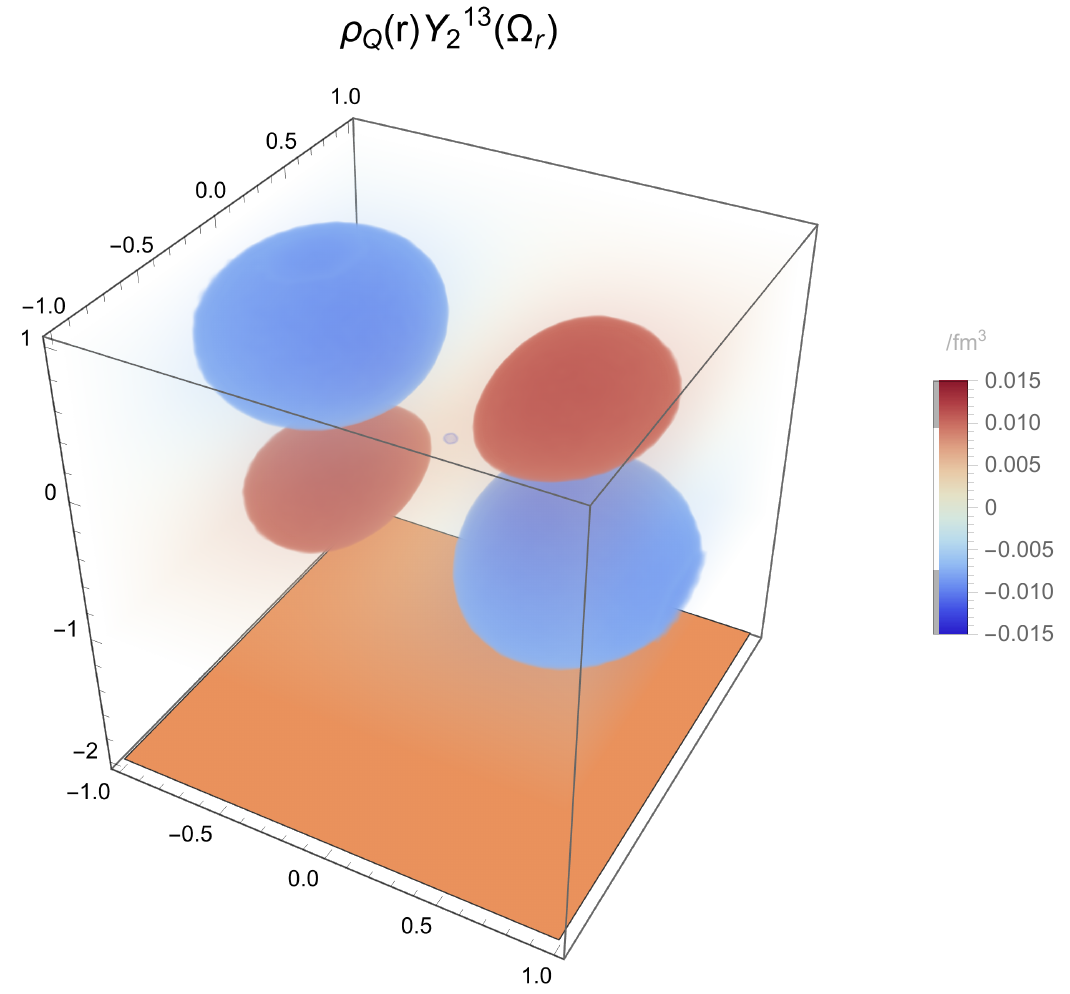}
\includegraphics[scale=0.545]{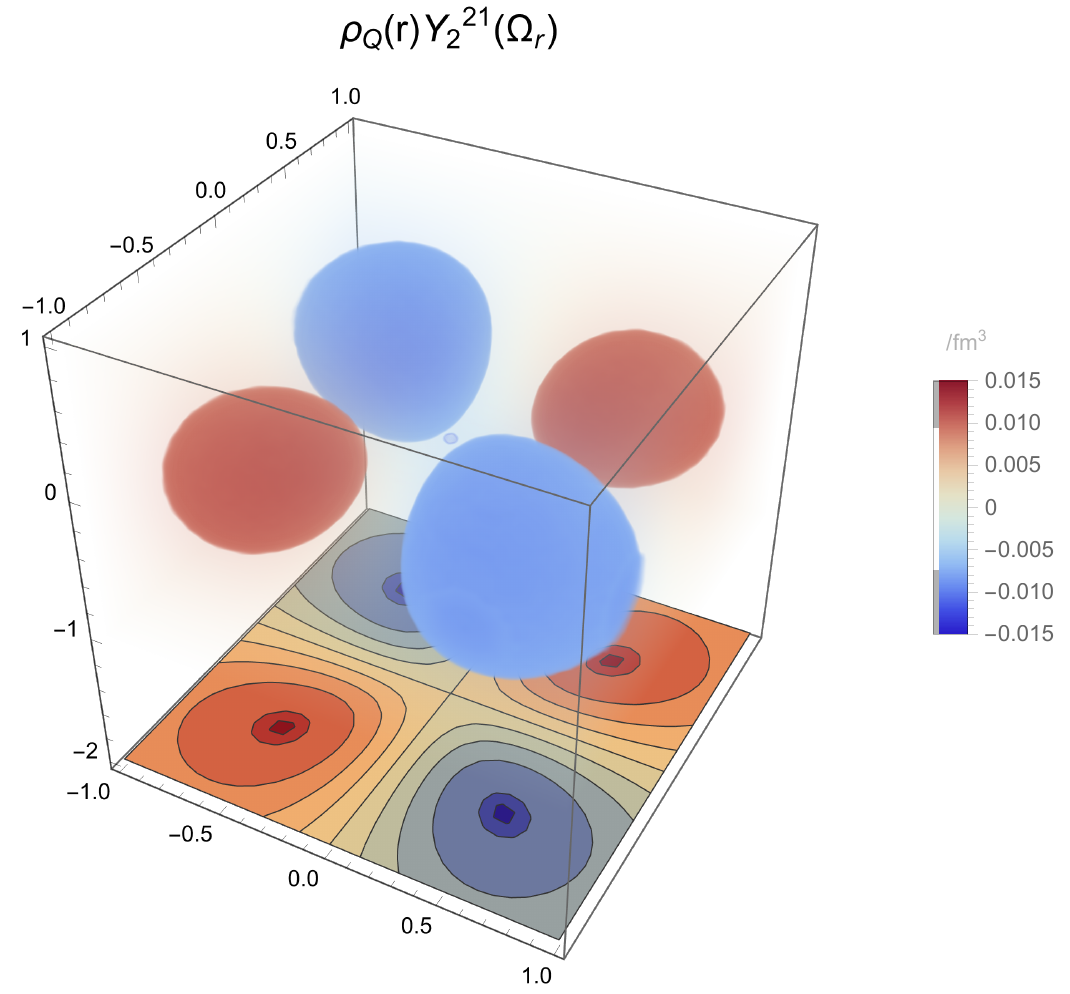}
\includegraphics[scale=0.545]{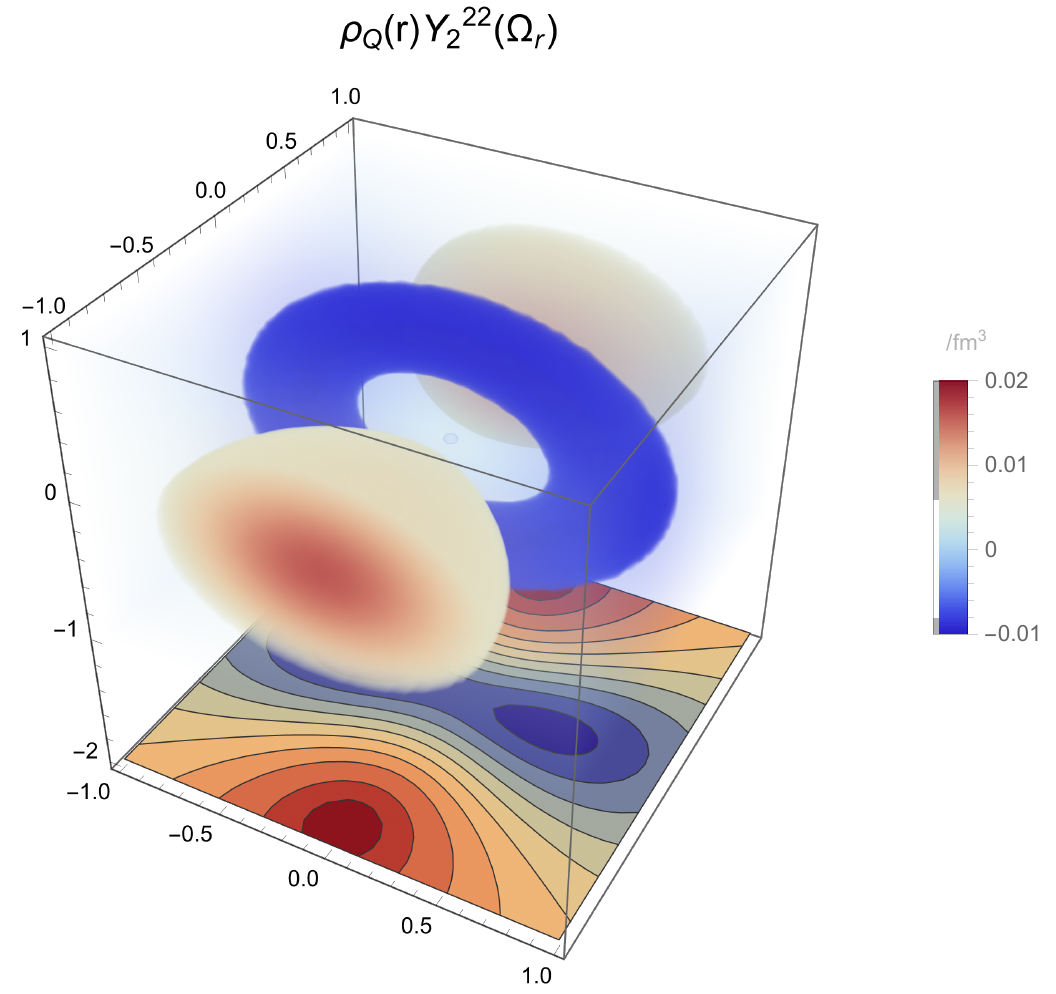}
\includegraphics[scale=0.545]{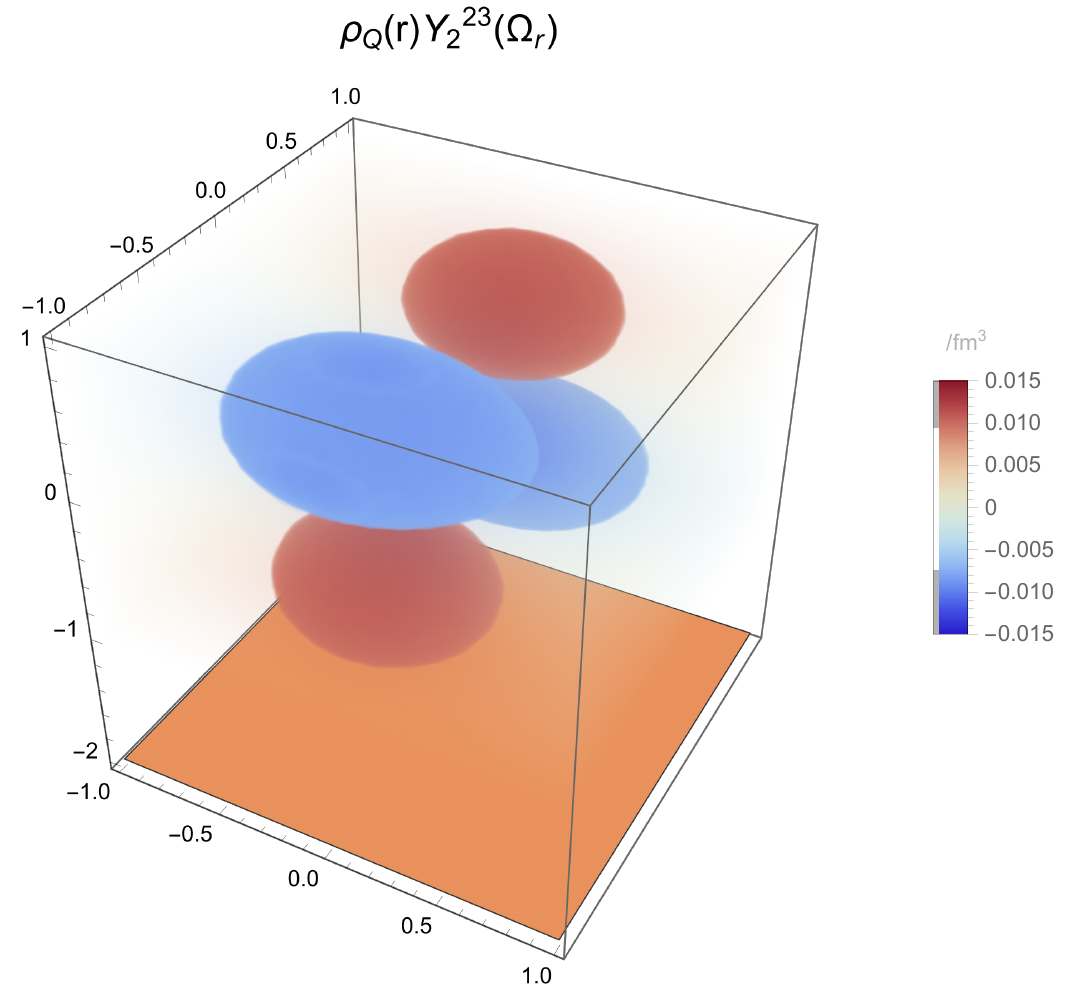}
\includegraphics[scale=0.545]{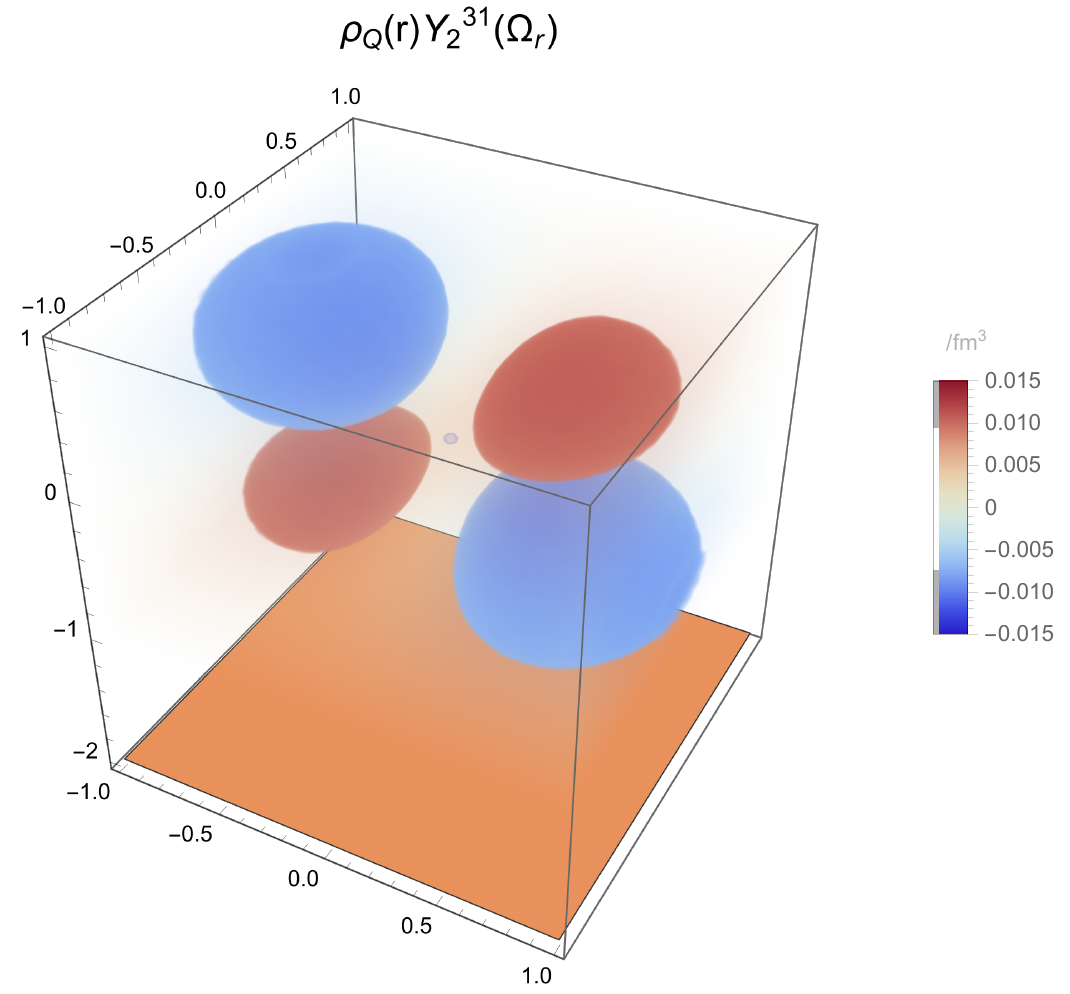}
\includegraphics[scale=0.545]{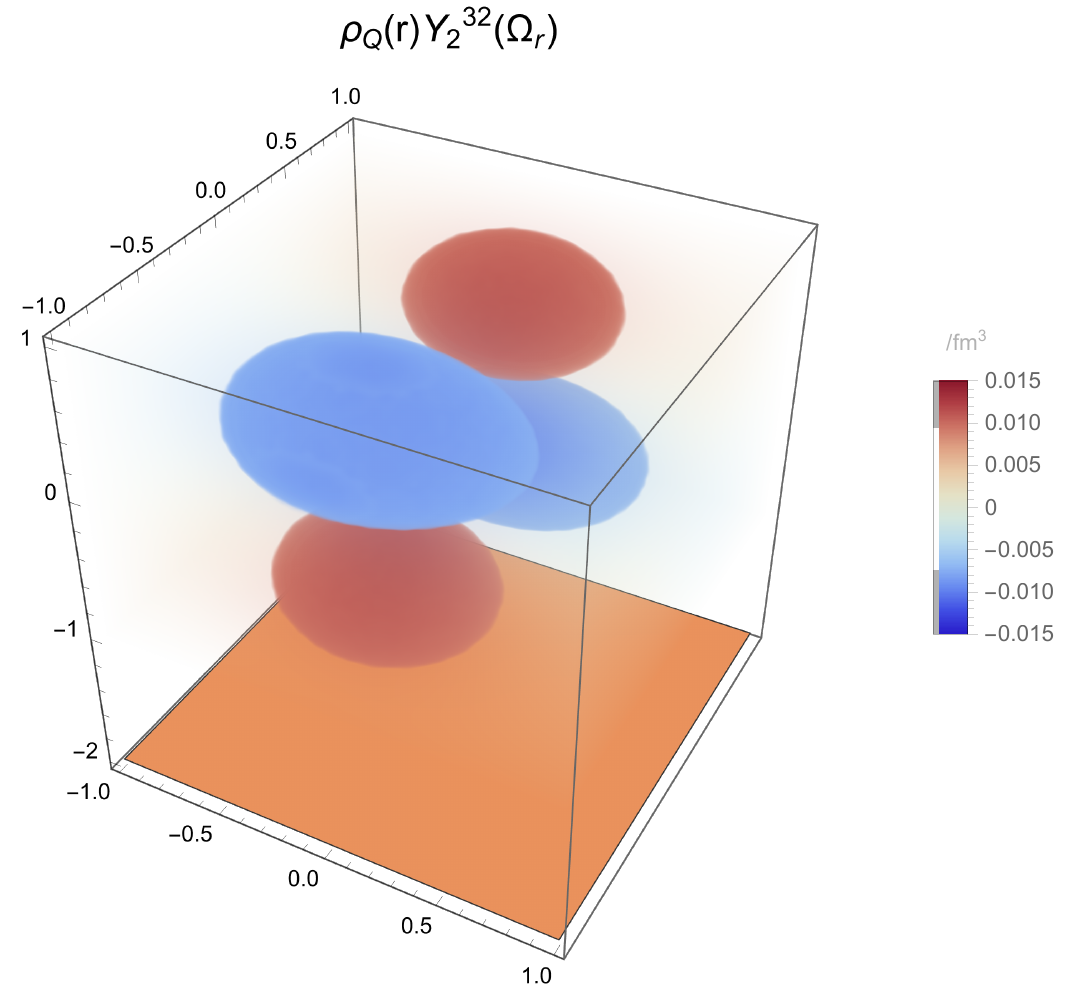}
\includegraphics[scale=0.545]{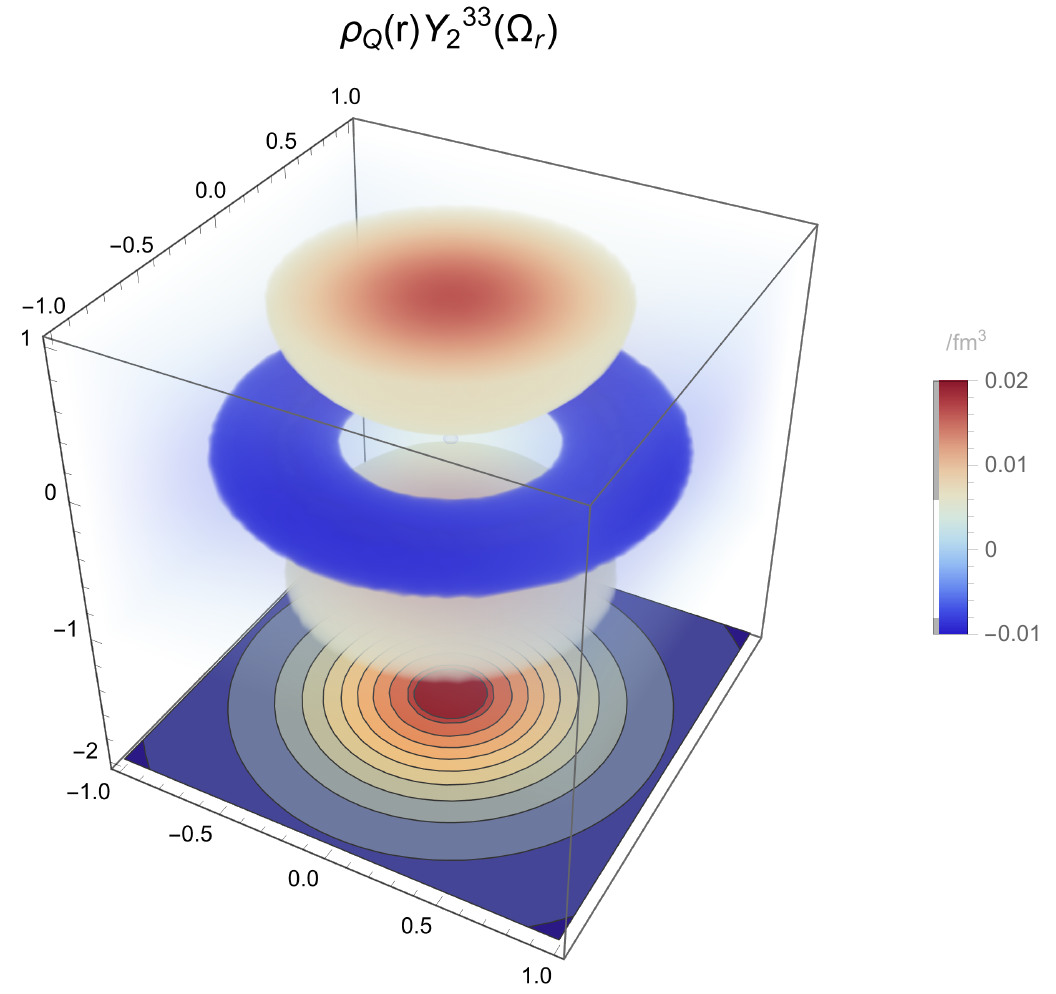}
\caption{Each component of the 3D electric quadrupole distribution $\rho_{Q}Y^{ij}$ is visualized, and the corresponding Abel image is drawn at the bottom plane of the box frame.}
\label{fig:2}
\end{figure}
To see them more clearly, in Fig.~\ref{fig:2}, we visualize each component of the quadrupole distribution $\rho_{Q}Y^{ij}$ in the 3D space and its Abel image on the bottom plane. Indeed, the off-diagonal components proportional to the single $z$ vanish. While the remaining the off-diagonal components $(i=1, j=2)$ and $(i=2, j=1)$ exhibit the quadrupole pattern, the diagonal components $(i=1, j=1)$ and $(i=2, j=2)$  are distorted due to the presence of both the monopole and quadrupole patterns. Interestingly, the component $(i=3, j=3)$ has only a monopole pattern, which means that it entirely flows into the charge distribution.  Therefore, the quadrupole patterns in the diagonal part together with those in the off-diagonal part constitute the 2D $2$-rank irreducible tensor. The remaining monopole patterns in the diagonal part differently affect the charge distribution $\rho^{\mathrm{(2D)}}_{C}$. As a result, the charge distribution is split into the $\rho^{\mathrm{(2D)}}_{C1}$ and $\rho^{\mathrm{(2D)}}_{C2}$.

\begin{figure}
\includegraphics[scale=0.275]{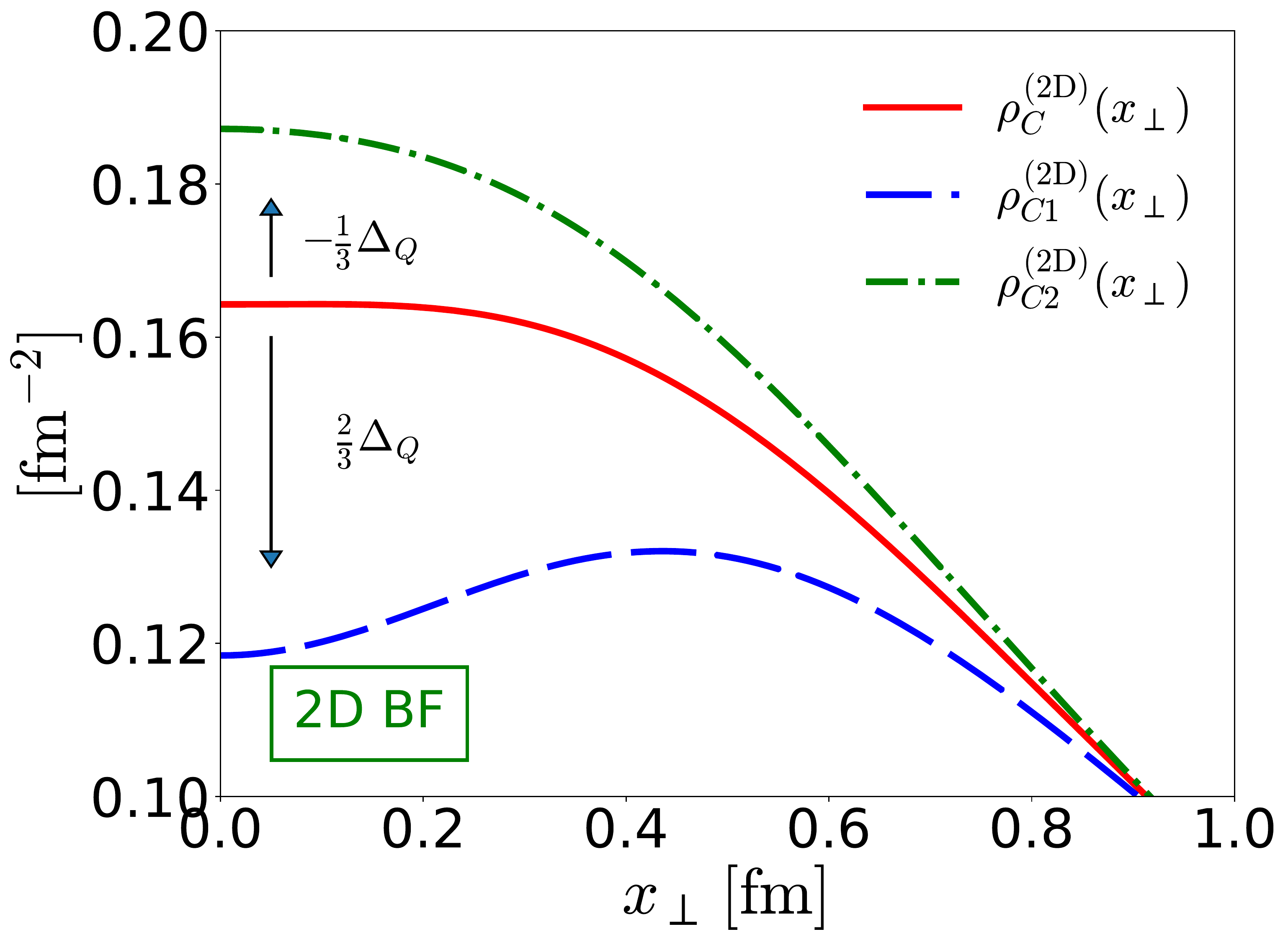}
\includegraphics[scale=0.275]{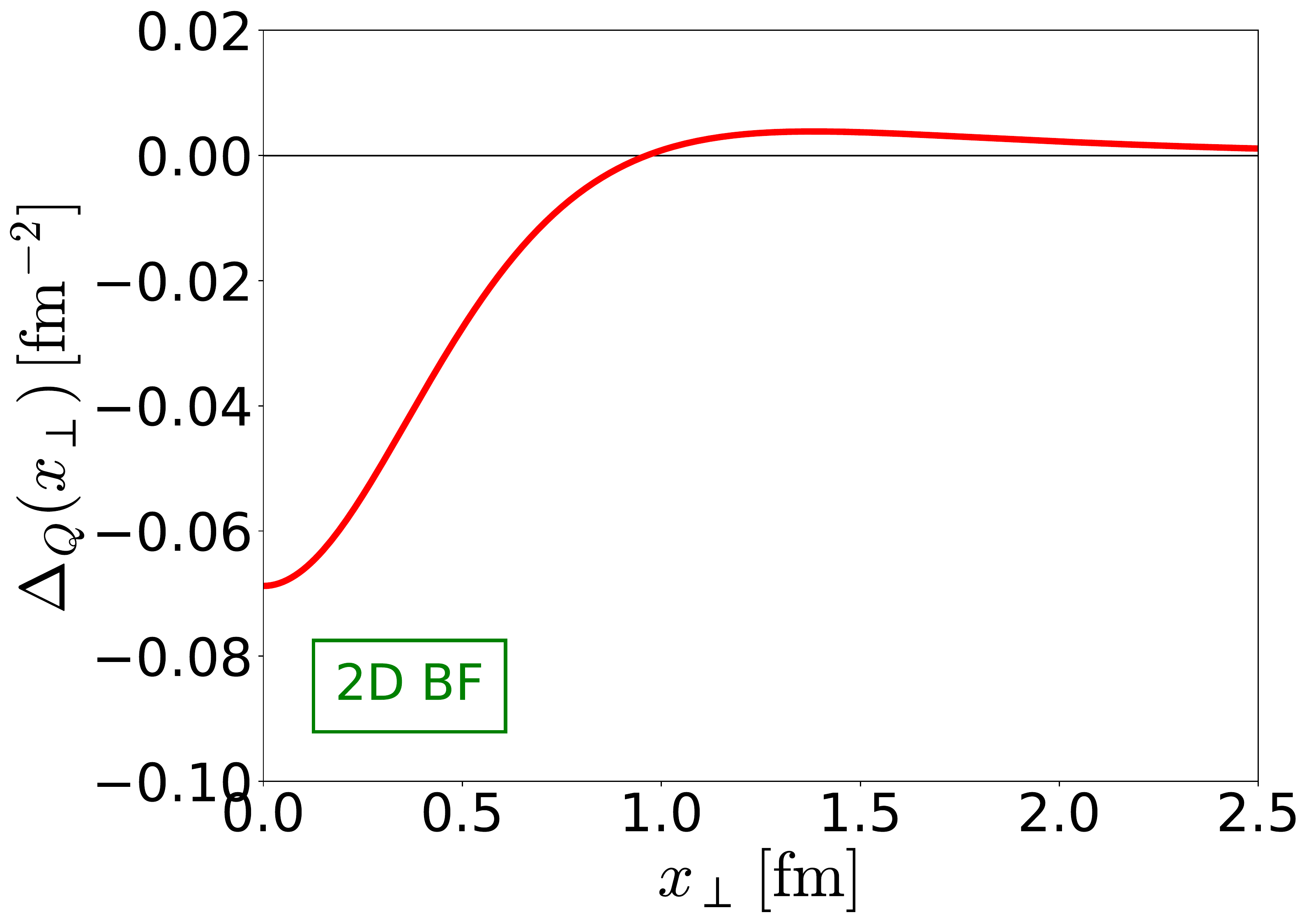}
\caption{In the left panel, we present the charge distributions $\rho^{\mathrm{(2D)}}_{C1}$ and $\rho^{\mathrm{(2D)}}_{C2}$ when the deuteron spin is longitudinally and transversely polarized to the $z$-axis, respectively. In the right panel, we present the induced monopole distribution $\Delta_{Q}(x_{\perp})$.}
\label{fig:3}
\end{figure}

In the left panel of Fig.~\ref{fig:3}, we present the split charge distributions of the deuteron, depending on its spin polarization. If the deuteron spin is longitudinally polarized to the $z$-axis, then its charge distribution decreases as much as $2\Delta_{Q}/3$. On the other hand, if it is transversely polarized to the $z$-axis, its charge distribution increase as much as $-\Delta_{Q}/3$. We thus naturally recover the fact that if a particle has a null quadrupole distribution, the charge distribution degenerates in the spin polarization.
We also present the induced monopole distribution $\Delta_{Q}$ in the right panel of Fig.~\ref{fig:3}. Since the contribution of the inner part of the nodal point cancels out that of the outer part, $\Delta_{Q}$ does not affect the normalization of the charge. It can be seen by the given obvious relation
\begin{align}
\int d^{2}x_{\perp} \, \Delta_{Q}(x_{\perp}) = \int d^{2}x_{\perp} \, \frac{\partial^{2}_{\mathrm{(2D)}}\tilde{G}^{\mathrm{(2D)}}_{Q}(x_{\perp})}{4m^{2}} =0,
\end{align}
However, this function contributes to the shape of the charge distribution and can be quantified by the charge radius. Interestingly, the difference between charge radii $\langle x^{2}_{\perp} \rangle^{\mathrm{(2D)}}_{C1}$ and $\langle x^{2}_{\perp} \rangle^{\mathrm{(2D)}}_{C2}$ is found to be the quadrupole moment of the deuteron $Q_{d}$:
\begin{align}
\langle x^{2}_{\perp} \rangle^{\mathrm{(2D)}}_{C1} - \langle x^{2}_{\perp} \rangle^{\mathrm{(2D)}}_{C2} = \int d^{2}x_{\perp} x^{2}_{\perp} \, \Delta_{Q}(x_{\perp})  = Q_{d} = 0.286 \, [\mathrm{fm}^{2}].
\end{align}
It indicates that the large value of the quadrupole moment is responsible for the sizable difference between the charge distributions $\rho^{\mathrm{(2D)}}_{C1}$ and $\rho^{\mathrm{(2D)}}_{C2}$.
\begin{figure}
\includegraphics[scale=0.275]{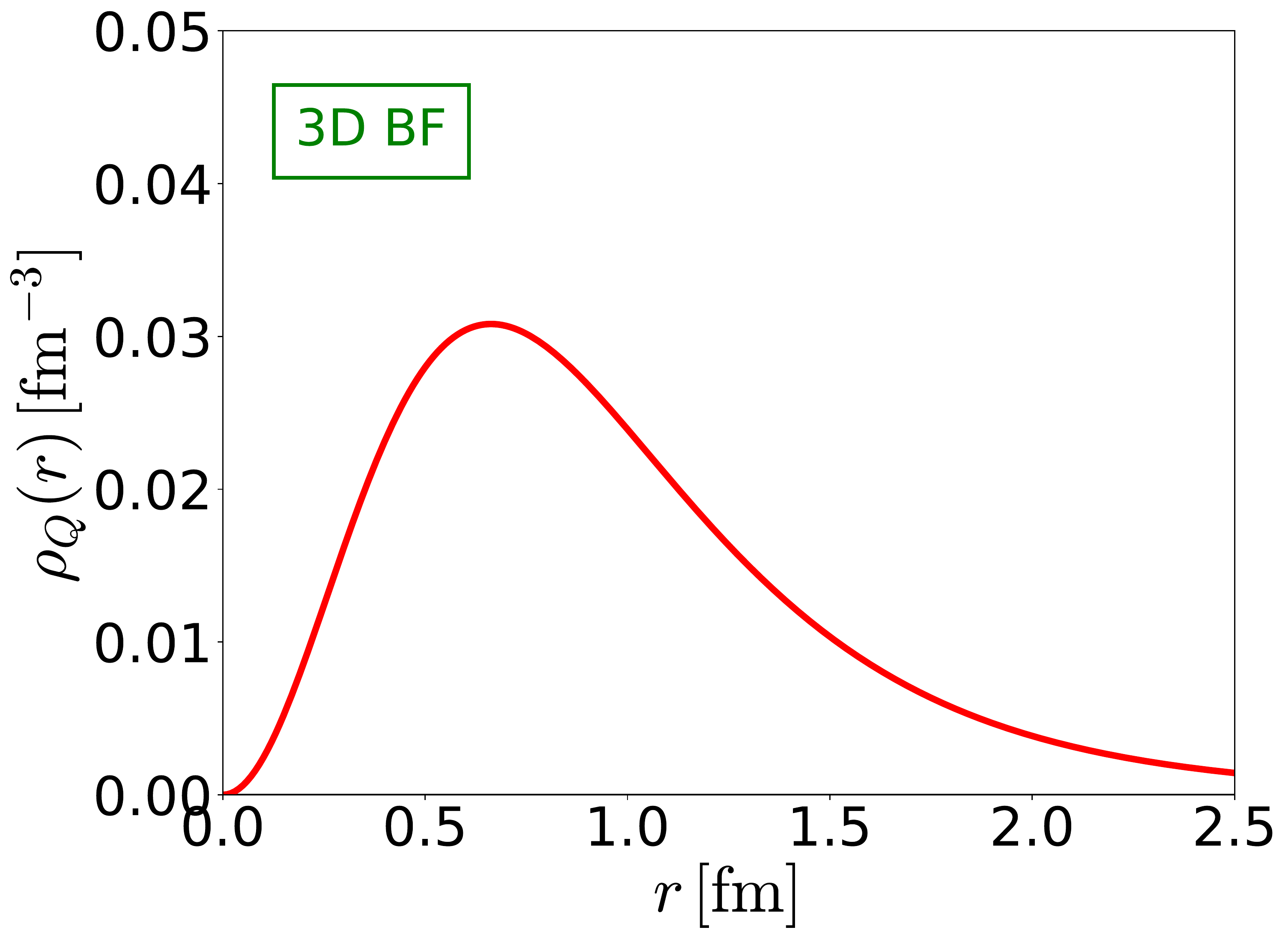}
\includegraphics[scale=0.275]{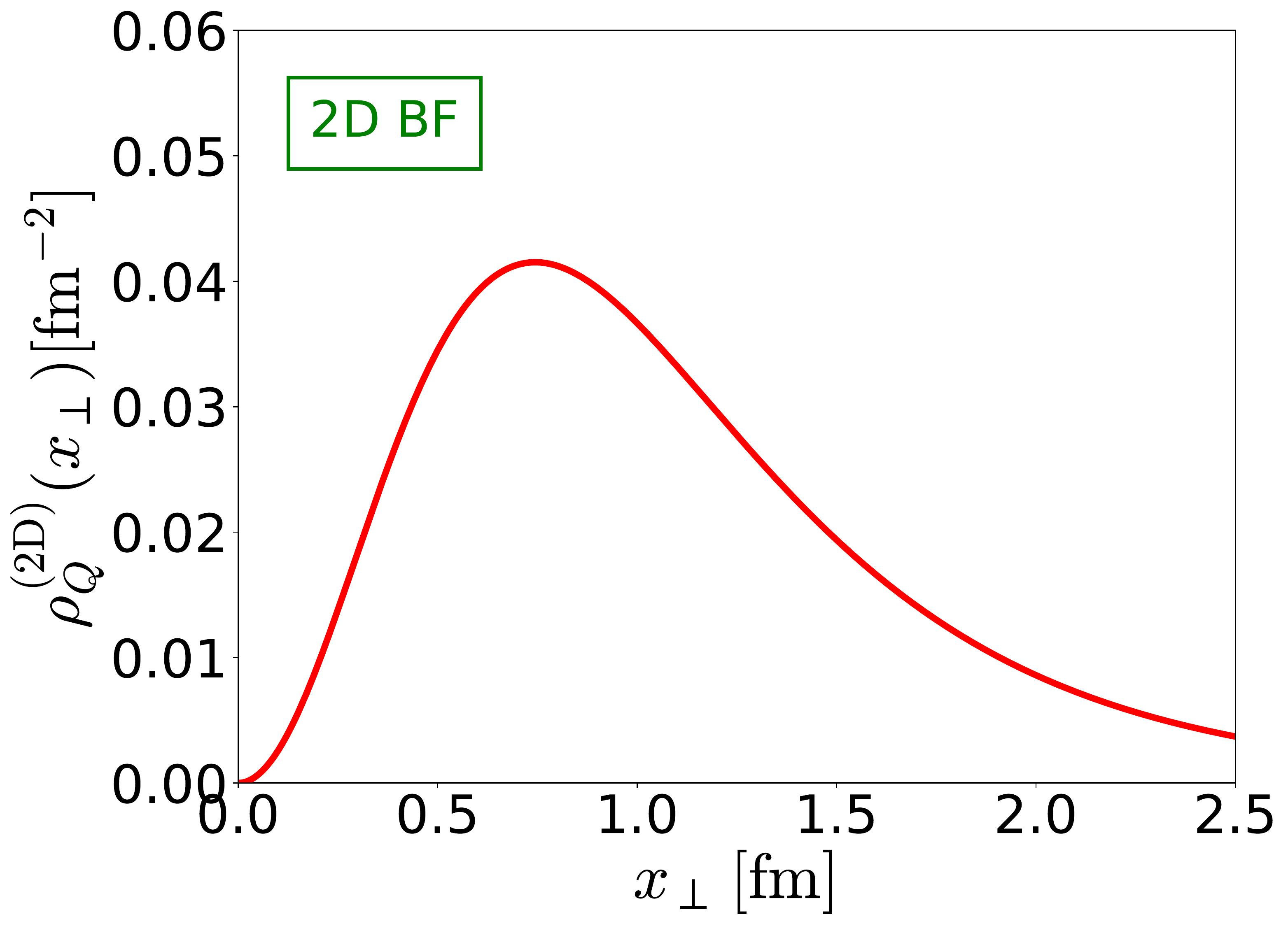}
\caption{In the left (right) panel, we present the electric quadrupole distribution $\rho_{Q}$ ($\rho^{\mathrm{(2D)}}_{Q}$) in the 3D BF (2D BF).}
\label{fig:4}
\end{figure}
In Fig.~\ref{fig:4}, we present quadrupole distributions in 3D BF $\rho_{Q}$ and 2D BF $\rho^{\mathrm{(2D)}}_{Q}$ for completeness. We find that the quadrupole distribution in the 2D BF gets concentrated on the center of the deuteron in comparison with that in the 3D BF.

\begin{figure}
\includegraphics[scale=0.275]{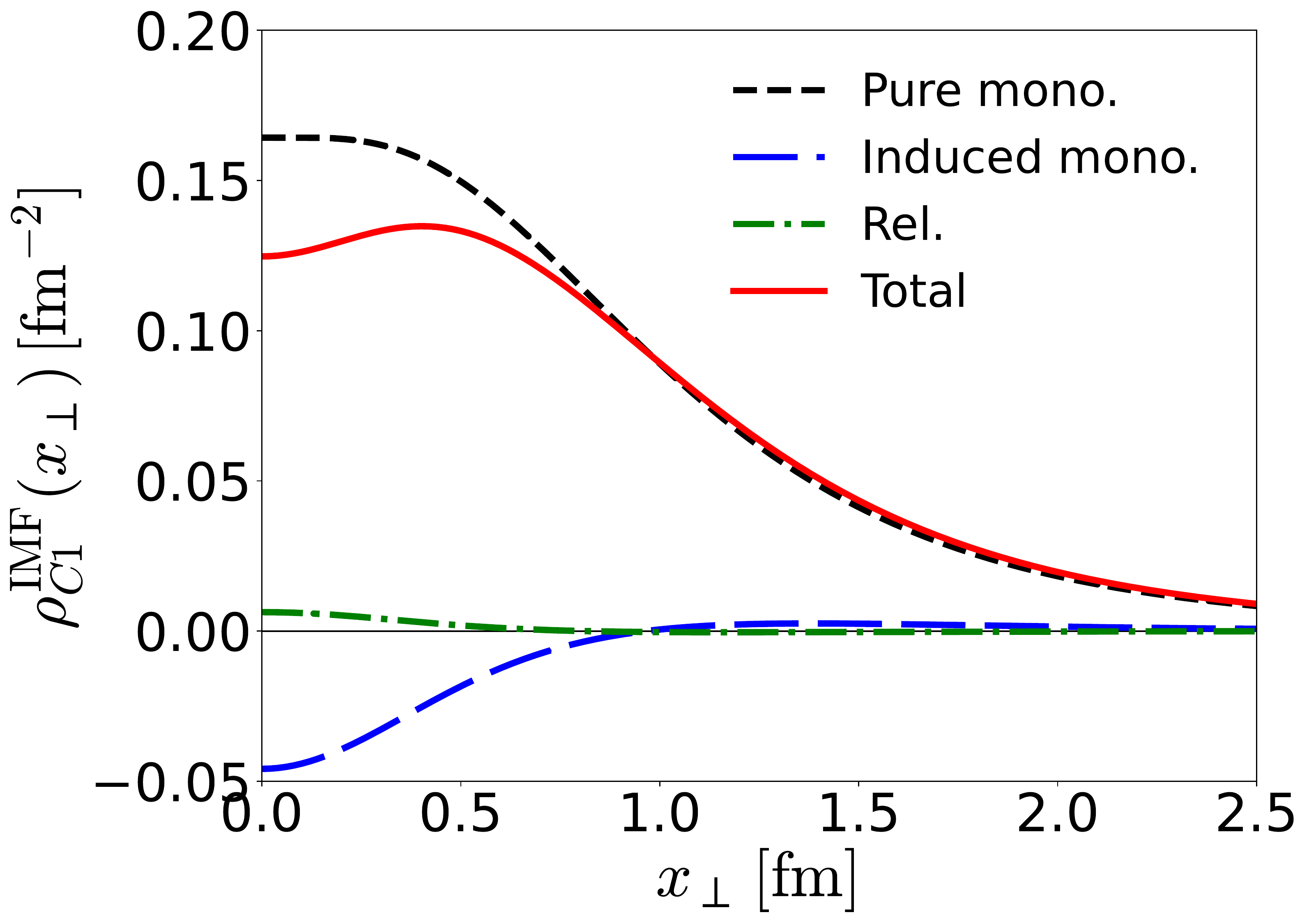}
\includegraphics[scale=0.275]{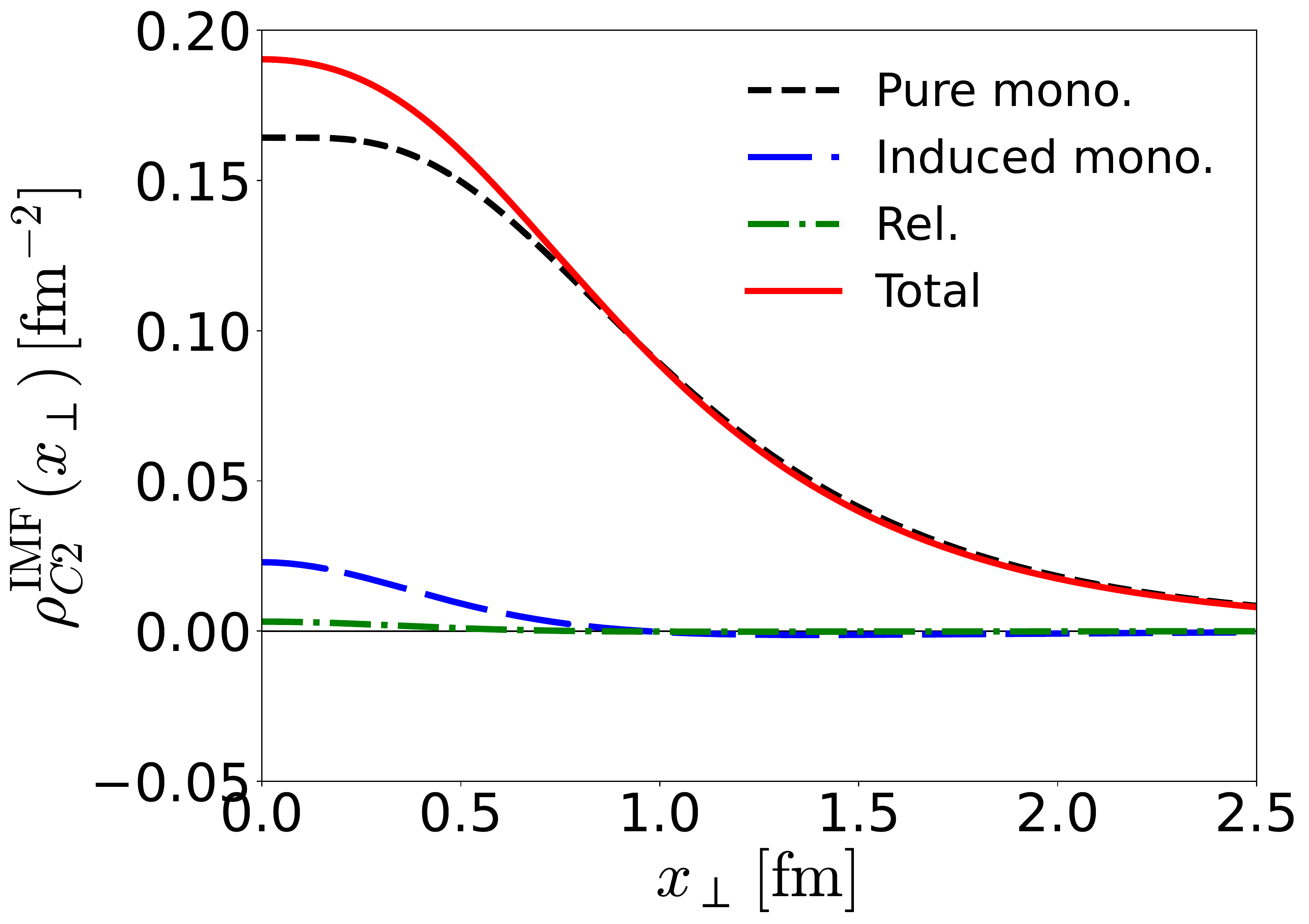}
\caption{The left~(right) panel presents the 2D charge distribution $\rho^{\mathrm{IMF}}_{C1}$~($\rho^{\mathrm{IMF}}_{C2}$) in the IMF for the deuteron. The long-dashed, dot-dashed, and short-dashed curves denote the induced monopole~(Induced mono.)~$\Delta_{Q}$,  relativistic~(Rel.)~$\tilde{G}_{W}$, and pure monopole~(Pure mono.)~$\rho^{\mathrm{(2D)}}_{C}$ contributions, respectively. The solid curve depicts the sum of the separate contributions. The EM form factors of the deuteron are taken from the parametrization given in Ref.~\cite{JLABt20:2000qyq}.}
\label{fig:5}
\end{figure}
When the deuteron is boosted to the IMF, its EM distributions are subjected to the relativistic corrections as well as the geometrical (or induced monopole) contribution $\Delta_{Q}$. In Fig.~\ref{fig:5}, we present the spin-dependent charge distributions $\rho^{\mathrm{IMF}}_{C1}$ and $\rho^{\mathrm{IMF}}_{C2}$ in the IMF. As shown in Eq.~\eqref{eq:decom}, they can be decomposed into the pure charge $\rho^{\mathrm{(2D)}}_{C}$, induced monopole $\Delta_{Q}$, and relativistic $\tilde{G}_{W}$ contributions. As discussed in the previous section, the leakage of the quadrupole distribution to the monopole one is solely responsible for the spin dependence of the charge distribution in the 2D BF. When the deuteron is boosted to the $z$-direction, the relativistic contributions come into play and differently contribute to charge distribution $\rho^{\mathrm{IMF}}_{C1}$ and $\rho^{\mathrm{IMF}}_{C2}$. This is the other origin of the spin dependence of the charge distribution. As shown in Fig.~\ref{fig:5}, while the deuteron is considerably affected by the induced monopole contribution $\Delta_{Q}$ due to the sizable quadrupole form factor, it is marginally subjected to relativistic effects $\tilde{G}_{W}$. It indicates that the main reason for the spin dependence of the charge distribution is attributed to the geometrical difference between 2D and 3D distributions. It should be distinguished from the Lorentz boost effects. This spin-dependent charge distribution will appear for any higher-spin particle such as $\rho$ meson, $\Delta$ baryon, and so on.
\begin{figure}
\includegraphics[scale=0.275]{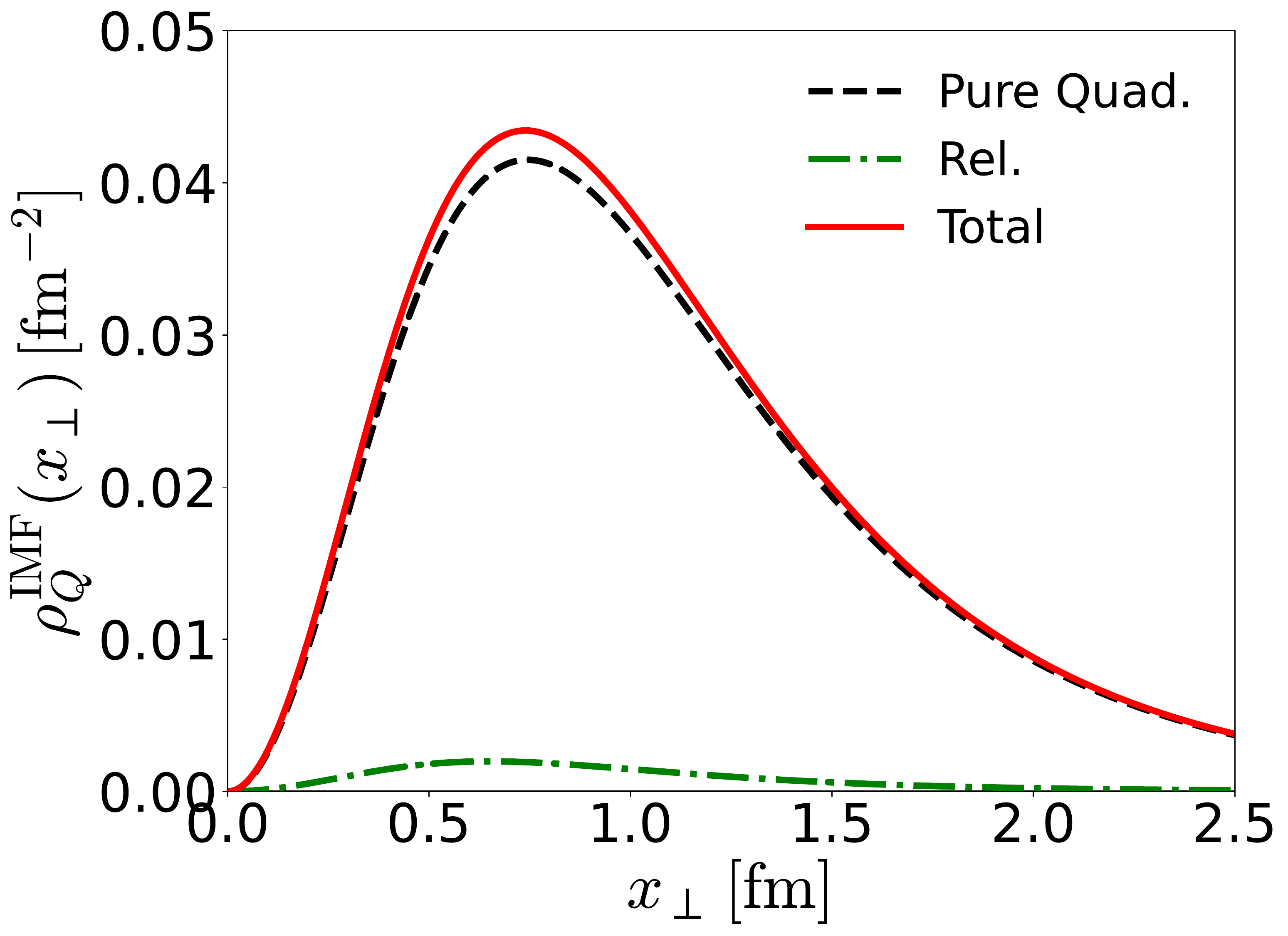}
\caption{The 2D electric quadrupole distribution $\rho^{\mathrm{IMF}}_{Q}$ for the deuteron in the IMF. The dot-dashed, and short-dashed curves denote the relativistic~(Rel.)~$\tilde{G}_{W}$, and pure quadrupole~(Pure quad.)~$\rho^{\mathrm{(2D)}}_{Q}$ contributions, respectively. The solid curve depicts the sum of the separate contributions. The EM form factors of the deuteron are taken from the parametrization given in Ref.~\cite{JLABt20:2000qyq}.}
\label{fig:6}
\end{figure}
In Fig.~\ref{fig:6}, we present the electric quadrupole distribution $\rho^{\mathrm{IMF}}_{Q}$. It is also decomposed into the relativistic $\tilde{G}_{W}$ and pure quadrupole $\rho^{\mathrm{(2D)}}_{Q}$ contributions. There is no induced multipole contribution and is a rather small relativistic contribution. To quantify the relativistic and induced monopole contributions we estimate the values of the charge radii and the electric quadrupole moment in Tab.~\ref{tab:1}. We found that the induced monopole contributions indeed dominate over the relativistic ones for the charge radii. We also found that the relativistic contributions to the electric quadrupole moment are negligible.
\begin{table}[htp]
\setlength{\tabcolsep}{5pt}
\renewcommand{\arraystretch}{1.5}
\caption{Separate pure multipole~(Pure mul.), induced monopole~(Induced mono.), and relativistic~(Rel.) contributions to the charge radii $\langle x^{2}_{\perp} \rangle^{\mathrm{IMF}}_{C1}$ and $\langle x^{2}_{\perp} \rangle^{\mathrm{IMF}}_{C2}$ and the electric quadrupole moment $Q^{\mathrm{IMF}}_{d}$ of the deuteron} 
\begin{tabular}{c | c c c } 
\hline
\hline{}
\centering
Deuteron & $ \langle x^{2}_{\perp} \rangle^{\mathrm{IMF}}_{C1}$ ($\mathrm{fm}^{2}$)  & $\langle x^{2}_{\perp} \rangle^{\mathrm{IMF}}_{C2}$ ($\mathrm{fm}^{2}$) & $Q_{d}^{\mathrm{IMF}}$ ($\mathrm{fm}^{2}$) \\
\hline
  Pure mul. & $2.90$ & $2.90$ & $0.286$ \\
  Induced mono. & $0.19$ & $-0.10$ & $-$\\
  Rel. & $-0.02$ & $-0.01$ & $0.008$ \\
  Total & $3.08$ & $2.80$ & $0.294$\\
\hline 
\hline
\end{tabular}
\label{tab:1}
\end{table}

\section{summary and conclusions}

In this work, we aimed at investigating how the charge distributions of a spin-one particle are related in the three different frames 3D Breit, 2D Breit, and 2D infinite momentum frames. Since, in the various Ref.~\cite{Carlson:2007xd, Miller:2007uy, Miller:2010nz, Carlson:2009ovh, Alexandrou:2009hs, Lorce:2022jyi}, the helicity-amplitude form factors have been used instead of the multipole form factors, it was rather difficult to grasp the physical meaning of each form factor. We thus provide the electromagnetic multipole form factors and the corresponding multipole distributions in the Wigner sense. In addition, while the spin-dependent charge distribution of a higher-spin particle has been observed in the infinite momentum frame, there was no relevant explanation for that. In this work, by employing the angle-dependent Abel transformation, we found that the geometrical difference between the 3D and 2D Breit frames brings about the spin dependence of the charge distributions. Specifically, if one projects the $2$-rank irreducible tensor in the 3D space to the 2D space, it is reduced to the $2$-rank and $0$-rank irreducible tensors in the 2D space. It indicates that the presence of the quadrupole structure causes the induced monopole distribution. It finally results in the spin-dependent monopole distribution, which is also true for the mass distributions of a higher-spin particle. Therefore, the unique  mass radius in the 2D space cannot be determined for the higher-spin particle~\cite{Freese:2019bhb, Pefkou:2021fni, Sun:2020wfo}. To have strict probabilistic distribution defined in the infinite momentum frame, we mapped the 2D charge distribution in the Breit frame to that in the infinite momentum frame through differential equations. They include the information on the Lorentz boost of the target. This Lorentz boost causes relativistic effects which consist of the contributions of the Wigner spin rotations and the mixture of the temporal and spatial components of the electromagnetic current. Interestingly, the Lorentz boost also differently contributes to respective spin-dependent charge distributions. It is the other origin of the split monopole distributions. To estimate the typical contributions of the induced monopole and Lorentz boost to the spin-dependent charge distributions, we employ the parametrization given in Ref.~\cite{JLABt20:2000qyq} for the electromagnetic form factors of the deuteron. We found that the induced monopole contributions to the spin-dependent charge distributions dominate over the relativistic ones. It indicates that the main reason for the spin dependence of the charge distribution for the deuteron is the geometrical difference between the 2D and 3D spaces in the presence of the quadrupole structure.

It is straightforward and interesting to formulate the energy-momentum tensor distributions of a higher-spin particle in both the 2D Breit and 2D infinite momentum frames in terms of the multipole expansion. Since the energy-momentum tensor has a far more complicated structure in comparison with the electromagnetic one, it is essential to classify them in terms of the multipole expansion and connect them to the 3D ones~\cite{Polyakov:2018rew, Polyakov:2019lbq, Cosyn:2019aio, Kim:2020lrs, Panteleeva:2020ejw}. Especially, the 2D quadrupole structure of the pressure and shear forces are expected to affect the monopole structure of them in the infinite momentum frame which may bring about the non-trivial stability conditions for a higher-spin particle. It is a distinctive feature, unlike the nucleon and the pion.

\begin{acknowledgments}
J.-Y. Kim is very grateful to Bao-Dong Sun and Hyun-Chul Kim for invaluable
discussion. J.-Y. Kim is supported by the Deutscher Akademischer Austauschdienst(DAAD) doctoral
scholarship. 
\end{acknowledgments}

\end{document}